%% file: main.tex
\documentclass[
aps,prd,
showpacs,twocolumn,notitlepage,
amssymb,amsmath,amsfonts,
nofootinbib,superscriptaddress,
floats,
amsmath,amssymb,
aps,
]{revtex4-2}

\usepackage{graphicx}
\usepackage{xcolor}
\usepackage[colorlinks=true]{hyperref}
\hypersetup{citecolor=cyan,linkcolor=magenta}
\usepackage{multirow,array}

\usepackage{amsmath,amssymb,siunitx}
\usepackage{hyperref}
\usepackage{rotating}
\usepackage{ulem}
\usepackage{array}
\usepackage{color}
\usepackage{multirow}
\usepackage{float}

\usepackage{comment}

\begin{document}

\newcommand{\kk}[1]{{\color{magenta} #1}}
\newcommand{\addms}[1]{{\color{red} #1}}
\newcommand{\wlh}[1]{{\textcolor{blue}{#1}}}

\newcommand{\kh}[1]{{\color{green} #1}}

\newcommand{\kenta}[1]{{\color{cyan} #1}}
\newcommand{\scc}[1]{{\color{orange} #1}}

\title{Black hole-neutron star mergers with massive neutron stars in numerical relativity}

%\author{
%  Luohan Wang$^{1,3}$, Shichuan Chen$^{1,3}$, Kota Hayashi$^{1,2}$, Kyohei %Kawaguchi$^{1,2,4}$, Kenta Kiuchi$^{1,2}$, Masaru Shibata$^{1,2}$~
%}
\author{Shichuan Chen}
\affiliation{Department of Astronomy, School of Physics, Peking University, Beijing 100871, China}
\affiliation{Kavli Institute for Astronomy and Astrophysics, Peking University, Beijing 100871, China}
\affiliation{Max Planck Institute for Gravitational Physics (Albert Einstein Institute),
Am M{\"u}hlenberg 1, Postdam-Golm 14476, Germany}

\author{Luohan Wang}
\affiliation{School of Physics, Peking University, Beijing 100871, China}
\affiliation{Max Planck Institute for Gravitational Physics (Albert Einstein Institute),
Am M{\"u}hlenberg 1, Postdam-Golm 14476, Germany}

\author{Kota Hayashi}
\affiliation{Max Planck Institute for Gravitational Physics (Albert Einstein Institute),
Am M{\"u}hlenberg 1, Postdam-Golm 14476, Germany}
\affiliation{Center for Gravitational Physics and Quantum Information, Yukawa Institute for Theoretical Physics,
  Kyoto University, Kyoto 606-8502, Japan}

\author{Kyohei Kawaguchi}
\affiliation{Max Planck Institute for Gravitational Physics (Albert Einstein Institute),
Am M{\"u}hlenberg 1, Postdam-Golm 14476, Germany}
\affiliation{Center for Gravitational Physics and Quantum Information, Yukawa Institute for Theoretical Physics,
  Kyoto University, Kyoto 606-8502, Japan}
\affiliation{Institute for Cosmic Ray Research, The University of Tokyo, 5-1-5 Kashiwanoha, Kashiwa, Chiba 277-8582, Japan}
%\date{\today}

\author{Kenta Kiuchi}
\affiliation{Max Planck Institute for Gravitational Physics (Albert Einstein Institute),
Am M{\"u}hlenberg 1, Postdam-Golm 14476, Germany}
\affiliation{Center for Gravitational Physics and Quantum Information, Yukawa Institute for Theoretical Physics,
  Kyoto University, Kyoto 606-8502, Japan}

\author{Masaru Shibata}
\affiliation{Max Planck Institute for Gravitational Physics (Albert Einstein Institute),
Am M{\"u}hlenberg 1, Postdam-Golm 14476, Germany}
\affiliation{Center for Gravitational Physics and Quantum Information, Yukawa Institute for Theoretical Physics,
  Kyoto University, Kyoto 606-8502, Japan}

\begin{abstract}

We study the merger of black hole-neutron star (BH-NS) binaries in numerical relativity, focusing on the properties of the remnant disk and the ejecta, varying the mass of compactness of the NS and the mass and spin of the BH. We find that within the precision of our numerical simulations, the remnant disk mass and ejecta mass normalized by the NS baryon mass ($\hat{M}_{\rm{rem}}$ and $\hat{M}_{\rm{eje}}$, respectively), and the cutoff frequency $f_{\rm{cut}}$ normalized by the initial total gravitational mass of the system at infinite separation approximately agree among the models with the same NS compactness $C_{\rm{NS}}=M_{\rm{NS}}/R_{\rm{NS}}$,  mass ratio $Q=M_{\rm{BH}}/M_{\rm{NS}}$, and dimensionless BH spin $\chi_{\rm{BH}}$ irrespective of the NS mass $M_{\rm{NS}}$ in the range of $1.092$--$1.691\,M_\odot$. This result shows that the merger outcome depends sensitively on $Q$, $\chi_{\rm BH}$, and $C_{\rm{NS}}$ but only weakly on $M_{\rm{NS}}$. This justifies the approach of studying the dependence of NS tidal disruptions on the NS compactness by fixing the NS mass but changing the EOS. We further perform simulations with massive NSs of  $M_{\rm{NS}}=1.8M_{\odot}$, and compare our results of $\hat{M}_{\rm{rem}}$ and $\hat{M}_{\rm{eje}}$ with those given by existing fitting formulas to test their robustness for more compact NSs. We find that the fitting formulas obtained in the previous studies are accurate within the numerical errors assumed, while our results also suggest that further improvement is possible by systematically performing more precise numerical simulations.
\end{abstract}
% Our results give important insight in discovering the systemic behavior of merger results and obtaining semianalytical models for $\hat{M}_{\rm{rem}}$ and $\hat{M}_{\rm{eje}}$. \kk{bit too abstract.}
\maketitle

\section{Introduction} \label{sec:intro}

The first gravitational wave (GW) detection for GW150914 from a binary black-hole (BBH) merger heralded the opening of the era of GW astronomy~\cite{abbott2016feb}. Subsequently, at the event of GW170817, the first binary neutron-star (BNS) merger was observed not only through GWs but also through signals of electromagnetic (EM) counterparts by diverse instruments all over the world~\cite{abbott2017oct1,abbott2017oct2,abbott2017oct3,Goldstein2017,LIGO_VIRGO_21505}. %\kenta{Kiuchi: It is better to cite more papers}
GW data analysis for this event gave a constraint on the tidal deformability of the NS which excludes very stiff equations of state (EOS)~\cite{abbott2017oct1,De2018,abbott2018,abbott2019jan,Narikawa2020}. The observation of short gamma-ray burst (GRB), GRB\,170817A, and the kilonova, AT2017gfo, indicated that the NS is involved in the merger, and it is suggested that a remnant formed after a BNS merger is likely to be the central engine of the GRB~\cite{abbott2017oct3,Goldstein2017,Savchenko2017}. The observation of the kilonova indicated that an $r$-process nucleosynthesis could occur in the BNS merger. All these facts suggest that future observations of GWs and EM counterparts will provide us valuable opportunities to deepen the knowledge of the mechanism of short GRBs, kilonovae, and $r$-process nucleosyntheses of heavy elements.

As in the BNS mergers, the BH-NS mergers can also, in principle, generate EM counterparts of GW such as kilonovae when a significant amount of matter is ejected in the tidal disruption of the NS. Such EM counterparts can give information about the NS EOS~\cite{tanaka2013dec,abbott2018} %\kenta{Kiuchi: "give information about" instead of "indicate"}
, the origin of heavy elements produced through $r$-process nucleosynthesis~\cite{lattimer1974,eichler1989,Freiburghaus1999,Drout2017,Pian2017}, and cast light on physics beyond the nuclear saturation density. However, although two events GW200105 and GW200115~\cite{abbott2021jun2} are reported to be the BH-NS merger in 2021, and in addition, GW190425 (with masses of 1.12-1.68 $M_{\odot}$ and 1.61-2.52 $M_{\odot}$) could possibly be a BH-NS merger \cite{Abbott2020mar,kyutoku2020feb}, no multi-messenger detection of BH-NS mergers has been confirmed yet. This lack of EM counterpart observation for the previous events is consistent with theoretical predictions because the mass ratios of these binaries are as high as 4--5 and the BH spins are likely to be zero or retrograde, and thus, tidal disruption is unlikely to occur~\cite{shibata2011aug}. However, for interpreting future events, in which tidal disruption of NSs may occur, it is crucial to prepare theoretical models that quantify the relation between the properties of the BH-NS merger and the observables.

%Later, the GW event GW190425 from the merger of binary compact objects with masses of 1.12-1.68 $M_{\odot}$ and 1.61-2.52 $M_{\odot}$ \cite{Abbott2020mar} has not been determined to be a BH-neutron sta (BH-NS) merger or not \cite{kyutoku2020feb}. Finally, two events GW200105 and GW200115 are reported to be the BH-NS merger in 2021, but still no associated electromagnetic counterpart was detected \cite{abbott2021jun2}. 

Numerical simulations of BH-NS mergers play an important role in understanding the tidal disruption and mass ejection processes. Since 2006~\cite{shibata2006dec,shibata2007may,shibata2008apr,etienne2008apr,duez2008nov}, many numerical simulations have been performed for this to study the dependence of the merger behavior on binary parameters, focusing on the characteristic results of tidal disruptions~\cite{yamamoto2008sep,etienne2009feb,shibata2009feb,foucart2011jan,kyutoku2010aug,kyutoku2015aug,kawaguchi2016jun,hayashi2021feb,foucart2019may,brege2018sep,foucart2013apr,foucart2012dec,foucart2012feb,lovelace2013jun,foucart2014jul,kawaguchi2015jul,Kyutoku:2021icp}. Additionally, it is revealed that the cutoff frequency, one of the notable features of the GW signal in tidal disruptions, encodes information on the NS EOS, especially when it is stiff~\cite{shibata2009feb,Ferrari2010,kyutoku2010aug,kyutoku2011sep}. More importantly, previous studies~\cite{kyutoku2015aug,foucart2012dec,krger2020feb,kawaguchi2016jun,foucart2018oct} show that the restmass of the remnant matter located outside the apparent horizon and ejecta mass normalized by the NS baryon mass ($\hat{M}_{\rm{rem}}$ and $\hat{M}_{\rm{eje}}$, respectively) as well as the cutoff frequency of GWs depend sensitively only on the mass ratio $Q$, the BH dimensionless spin $\chi_{\rm BH}$, the compactness of NSs $C_{\rm{NS}}$ but not on the NS mass $M_{\rm{NS}}$. Based on this approximate parameter dependence, previous studies~\cite{foucart2012dec,foucart2018oct,kawaguchi2016jun,krger2020feb} of BH-NS mergers give some fitting formulas for the remnant mass $M_{\rm{rem}}$ and the ejecta mass $M_{\rm{eje}}$. These semi-analytical models for the merger results are valuable because numerical simulations of BH-NS mergers are resource-intensive. These fitting formulas for the remnant disk mass and ejecta mass are used to assess whether the EM counterparts would be present~\cite{Pannarale2014,foucart2012dec}, and constrain the binary parameters after the observation of EM signals~\cite{Radice2018,Hinderer2019,Coughlin2019,Barbieri2019,Andreoni2020,Ascenzi2019}. 

%Quantities such as the remnant disk mass, the cutoff frequency are useful in studying the criterion for tidal disruptions, and properties of the ejecta such as mass and velocity can also provide valuable information about the merger event. 

However, the typical range of NS mass in these previous simulations is limited to $M_{\rm{NS}}\approx 1.2$--$1.5M_{\odot}$~\cite{Tauris2017,Farrow2019}, with NS compactness between 0.12 and 0.19, although more massive NSs at least up to $\sim 2M_{\odot}$ exist in nature~\cite{demorest2010oct,Antoniadis2013}. It is not clear whether the fitting formulas obtained by the previous studies also hold quantitatively for more massive and compact NSs because they are not tested for more massive NSs. Given possible GW detections of such massive NSs in BH-NS binaries~\cite{Abbott2020mar,abbott2020jun}, quantitative investigation of the BH-NS mergers in such a parameter space is essential.  %Semianalytical models for the merger results are valuable because numerical simulations of BH-NS mergers are resource-intensive. Previous simulations \cite{foucart2012dec,foucart2018oct,kawaguchi2016jun,krger2020feb} of BH-NS mergers give some fitting formulas for the remnant mass $M_{\rm{rem}}$, the ejecta mass $M_{\rm{eje}}$, the cutoff frequency and the ejecta velocity, with NSs compactness $C_{\rm{NS}}\geq 0.18$ used in the simulations.

In this paper, we study BH-NS mergers in numerical relativity varying the NS mass and NS compactness for a range wider than in the previous studies. We first focus on BH-NS binaries with the several fixed combinations of mass ratio $Q$, BH dimensionless spin $\chi_{\rm BH}$, and NS compactness $C_{\rm{NS}}$ but with various NS masses $M_{\rm NS}$ in order to clarify how $\hat{M}_{\rm{rem}}$, $\hat{M}_{\rm{eje}}$, and $f_{\rm {cut}}m_0$~(where $m_0=M_\mathrm{BH}+M_\mathrm{NS}$) are sensitive to the NS mass $M_{\rm NS}$. We then perform simulations for a large value of the NS mass ($M_{\rm NS}=1.8\, M_\odot$), and examine the accuracy of the previous fitting formulas of $\hat{M}_{\rm{rem}}$, $\hat{M}_{\rm{eje}}$, and ejecta velocity $v_{\rm{eje}}$ for the system with a large value of the NS compactness ($C_{\rm NS}\ge 0.19$). We will show that within the range of the numerical accuracy, $\hat{M}_{\rm{rem}}$, $\hat{M}_{\rm{eje}}$, and $f_{\rm{cut}}m_0$ can be derived from the fitting formulas if the values of $\chi_{\rm{BH}}$, $Q$, and $C_{\rm{NS}}$ are identical: The fitting formulas previously obtained in  Refs.~\cite{foucart2012dec,foucart2018oct,krger2020feb} still work well for larger values of $C_{\rm{NS}}$ above $0.19$.

%Numerical simulations of BH-NS mergers are resource-intensive, which makes semianalytical models for the merger results very valuable. Previous simulations \cite{foucart2012dec,foucart2018oct,kawaguchi2016jun,krger2020feb} of BH-NS mergers give some fitting formulas for the remnant mass $M_{\rm{rem}}$ and the ejecta mass $M_{\rm{eje}}$, with less compact NSs ($C_{\rm{NS}}\geq 0.18$) used in the simulations.
%In this study, we are interested in whether these fitting formulas still perform well for larger $C_{\rm{NS}}$ above 0.19 and evaluate their validity analytically.

The paper is organized as follows. In Sec.~\ref{sec:methods} we briefly summarize numerical methods used in our study. Numerical results of simulations are presented in Sec.~\ref{sec:results}, focusing on the remnant disk mass, the ejecta mass, and the gravitational waveforms. In Secs.~\ref{sec:fixed-C} and~\ref{sec:mNS}, we present the results for the simulations with the fixed NS compactness and those for massive and compact NSs, respectively. Sec.~\ref{sec:conclusion} is devoted to a summary and discussion. Throughout this paper, all the quantities in tables and figures are shown in units of $c=G=M_{\odot}=1$ unless otherwise stated. Our convention of notation is summarized in Table~\ref{tab:quantity}.
\input{table_physical_quantities}

\section{Numerical methods}\label{sec:methods}

This section describes the numerical methods used in our study. Sec.~\ref{sec:simulation_methods} summarizes our numerical simulation methods, Sec.~\ref{sec:NS_EOS} describes the NS EOS, Sec.~\ref{sec:diagnostics} presents the diagnostics of simulations, Sec.~\ref{sec:fitting} presents the fitting formulas given from previous studies, and Sec.~\ref{sec:models} presents the models employed in this study.

\subsection{Simulation methods} \label{sec:simulation_methods}

We carry out numerical simulations using the SACRA-MPI code \cite{kiuchi2017oct}, which uses an adaptive-mesh-refinement (AMR) algorithm~\cite{yamamoto2008sep} and MPI/OpenMP hybrid parallelization to speed up the computation \cite{kiuchi2017oct}. SACRA solves the Einstein equation in a moving puncture version \cite{campanelli2006mar,baker2006mar,marronetti2008mar} of the Baumgarte-Shapiro-Shibata-Nakamura (BSSN) formulation \cite{shibata1995nov,baumgarte1998dec}, incorporating a Z4c constraint-propagation prescription locally \cite{hilditch2013oct}. 
Together with the Einstein equation, we solve hydrodynamics equations using the Harten-Lax-van Leer contact (HLLC) solver~\cite{mignone2005,white2016,kiuchi2022} in this paper. More details of the formulation, the gauge conditions, and the numerical scheme are described in Refs.~\cite{kyutoku2010aug,kyutoku2011sep,kyutoku2015aug,kawaguchi2015jul}. For actual computation, we employ the public spectral code FUKA (Frankfurt University Kadath)~\cite{Papenfort:2011} to generate initial data. We do not take into account magnetohydrodynamics and neutrino-transfer effects, since we focus on the evolution of the system up to 30\,ms after the merger for which these effects are not very important. 
%\kenta{Kiuchi: Strictly speaking, the initial data are generated by FUKA, not by LORENE. I would suggest to add the sentence "We employ the public spectral code FUKA (Frankfurt University Kadath)~\cite{Papenfort:2011} to generate initial data."}

The grid structure of SACRA is summarized as follows. Computational domains are composed of nested equidistant Cartesian grids, and each grid has $(2N, 2N, N)$ points in $(x, y,z)$ directions. 
We employ a cell-centered grid structure and the $x$-coordinate at $j$th grid point is given by $x_j=(j+1/2)\Delta x$, where $\Delta x$ is the grid resolution.
%\kenta{Kiuchi: Question. Do you employ the vertex center? If so, please add the discussion about how good the baryon mass conservation is somewhere.}
The equatorial-plane symmetry is imposed on the $z=0$ plane. We adopt $N=82$ as a fiducial value, with which the NS radius is covered by $\approx 66$ points in the finest grid. We also perform simulations with $N=62$ and $N=102$ for selected models to evaluate the numerical error and check the convergence of the numerical results (see Appendix~\ref{sec:appendix1}). 
The physical quantities in tables and figures, unless specified, are taken from results with $N=82$. In our study, the inspiral motion before the merger is followed at least for 5 orbits.

We prepare 10 refinement levels for the AMR computational domains. 6 coarser domains ($l_c$) cover both BH and NS with the origins fixed at the center of the mass of the binary system. Two sets of 4 finer domains ($l_f$) comove with either BH or NS, covering the region of their vicinity. Namely, we prepare $2l_f+l_c$ computational grid domains spanning $l_c+l_f$ refinement levels. Starting from the coarsest level as $l=0$, the $l$th level has a grid spacing, $\Delta x_l=L/(2^l N)$, and at the finest level, $\Delta x=L/(2^{l_f+l_c-1}N)$, where $L$ is the size of the computational domain, which covers $[-L:L]\times[-L:L]\times[0:L]$ for $x$-$y$-$z$.

%\subsection{Zero-temperature equation of state}\label{sec:NS_EOS}
\subsection{Equation of state}\label{sec:NS_EOS}

Since the lifetime of NS binaries is typically much longer than the cooling time of NSs, NSs in the late inspiral stage are believed to be cold enough to be modeled by zero-temperature EOS~\cite{lattimer2004apr}. Thus, for modeling the NSs prior to the merger, we use the zero-temperature EOS in our simulations. In this paper, we model such EOSs by piecewise polytropes~\cite{read2009jun,read2009jun2,lackey2012feb}, which are written in the form: 
\begin{equation}
    P(\rho)=\kappa_i\rho^{\Gamma_i}\ (\rho_i\leq \rho\leq \rho_{i+1}),
\end{equation}
where $\kappa_i$ are constants and other quantities are described in Table~\ref{tab:quantity}. We perform simulations for models with both two pieces and four pieces, with $i\in \left\{0, 1\right\}$ and $i\in \left\{0, 1,2, 3\right\}$ respectively. $\rho_0=0$, and for $\rho\leq \rho_1$, we adopt $\Gamma_0=1.35692395$ and $\kappa_0/c^2=3.99873692\times10^{-8}\ \rm{g^{1-\Gamma_0}cm^{3\Gamma_0-3}}$~\cite{read2009jun}. The boundary condition at $\rho_i$ is $\kappa_i \rho_i^{\Gamma_i}=\kappa_{i+1}\rho_i^{\Gamma_{i+1}}$, which requires the continuity of pressure at the interface between $i^{\rm th}$ and $(i+1)^{\rm th}$ region. 

For different EOS models, we take different fiducial pressures $P_{\rm{fid}}$ at the fiducial density $\rho_{\rm{fid}}=10^{14.7}\ \rm{g\,cm^{-3}}$. For piecewise polytropes with two pieces, we take $\rho_2\rightarrow+\infty$, $\Gamma_1=3.000$, and $\kappa_1$ is defined by  $\kappa_1=P_{\rm{fid}}/\rho_{\rm{fid}}^{\Gamma_1}$. For piecewise polytropes with four pieces, we take $\rho_2=\rho_{\rm{fid}}$, $\rho_3=10^{15}~{\rm g\,cm^{-3}}$, $\rho_4\rightarrow+\infty$, and $\kappa_2$ is defined by $\kappa_2=P_{\rm{fid}}/\rho_{\rm{fid}}^{\Gamma_2}$.
% \kenta{Kiuchi: $\rho_3=10^{15}~{\rm g/cm^3}$, and $\rho_4=\infty$}. 
Other parameters for different EOS models and NS initial properties related to the EOS are listed in Table~\ref{tab:EOS}. 

In numerical simulations, we add the thermal part of the EOS to the zero-temperature part described above. Our implementation for it is the same as that described in Refs.~\cite{hotokezaka2013} with an adiabatic index of the thermal part $\Gamma_{\rm th}=1.8$.
%\kenta{Kiuchi: I prefer to mention the value of $\Gamma_\rm{th}$ explicitly.}

\input{table_EOS}

\subsection{Diagnostics}\label{sec:diagnostics}

\subsubsection{Remnant disk and ejecta}

After the merger, the fate of the matter originating from the NS can be divided into three types. The matter directly falls into the BH, forms a remnant disk, and becomes unbound from the system, i.e. ejecta. Evaluating the properties of the disk and the ejecta is essential for discussing resultant models of EM counterparts. Here we describe the method to evaluate these quantities. In the following, we will define the time of merger $t_{\rm merger}$ as the time when $0.01 M_{\odot}$ of the NS matter falls into the apparent horizon.

At each time slice, we evaluate the remnant disk by the rest mass outside the apparent horizon, i.e. the remnant disk mass with the integral
\begin{equation}
    M_{\rm >AH}:=\int_{r>r_{\rm AH}}\rho_* \,d^3x
    \label{eq:Mrem}
\end{equation}
where $r_{\rm{AH}}=r_{\rm{AH}}(\theta,\phi)$ denotes the coordinate radius of the apparent horizon and $\rho_*$ is denfined in Table~\ref{tab:quantity}. 

The unbound matter of the system, ejecta, is defined as the matter satisfying $-u_t>1$, because in this paper we focus only on the dynamical ejecta, for which the thermal effect is minor. 
% Therefore, the total energy of the ejecta is
% \begin{equation}\label{eq:total_energy}
%      E_{\rm eje}:=\int_{-u_t>1,r>r_{\rm AH}}\rho_*(-u_t)\  \rm{d}^3x.
% \end{equation}
The mass of ejecta at each time slice is defined by the integral
\begin{equation}
    M_{\rm eje}:=\int_{-u_t>1,r>r_{\rm AH}}\rho_*\,d^3x.
    \label{eq:Meje}
\end{equation}

% The internal energy of the ejecta is
% \begin{equation}\label{eq:internal}
%     U_{\rm{eje}}:=\int_{-u_t>1,r>r_{\rm AH}}\rho_*\epsilon\  \rm{d}^3x.
% \end{equation}

Assuming that the thermal energy of the ejecta is much smaller than the kinetic energy, the asymptotic kinetic energy of the ejecta, $T_{\rm eje}$, can be defined by
\begin{equation}
    T_{\rm eje}:=\int_{-u_t>1,r>r_{\rm AH}}\rho_* (-u_t)\,d^3x. 
    \label{eq:T_eje}
\end{equation}
Then the average velocity of the ejecta is estimated from the asymptotic kinetic energy $T_{\rm{eje}}$, and the ejecta mass $M_{\rm eje}$, as
\begin{equation}
    v_{\rm eje}:=\sqrt{\frac{2T_{\rm eje}}{M_{\rm eje}}}.
    \label{eq:v_eje}
\end{equation}
%\sout{This estimation of the ejecta velocity does not take the gravitational binding energy associated with remnant BH-disk systems into account.}
%\kk{(!!!Again, please check if this formula is really used for the ejecta velocity estimation!!!)} \kk{(!!!I think this is not true. Please ask Kota.!!!)}
%Therefore, we typically measure the quantities of the ejecta at 10 ms after the onset of the merger, when the dominant portion of the ejecta has come to distant regions but still resides in our computational domains.}\kk{(Using Eq.~\ref{eq:T_eje}, the correction of the gravitational binding energy is automatically taken into account.)}
Following the previous studies~\cite{kyutoku2011sep,kyutoku2015aug, kawaguchi2016jun,foucart2012dec,foucart2018oct,krger2020feb}, we evaluate all these ejecta quantities at 10\,ms after the onset of the merger.

\subsubsection{Black hole parameters}

During the inspiral or post-merger stage, the parameters of BHs are estimated from the quantities of the apparent horizons. If we assume that the spacetime near the apparent horizon is stationary, the equatorial circumferential radius $C_{\rm{e}}$ and the area of the apparent horizon $A_{\rm{AH}}$ can be approximated as~\cite{shibata2009feb}
\begin{equation}
    C_{\rm e}=4\pi M_{\rm BH},
    \label{eq:C_e}
\end{equation}
and
\begin{equation}
    A_{\rm AH}=8\pi M_{\rm BH}^2\left(1+\sqrt{1-\chi_{\rm BH}^2}\right).
    \label{eq:A_AH}
\end{equation}
Then BH mass $M_{\rm BH}$ and dimensionless spin parameter $\chi_\mathrm{BH}$
can be evaluated from Eqs.~(\ref{eq:C_e}) and (\ref{eq:A_AH}).

% \begin{equation}
%     M_{\rm BH}=\frac{C_{\rm e}}{4\pi}
%     \label{eq:M_BH}
% \end{equation}

% \begin{equation}
%     \chi_{\rm{BH}}=\frac{1}{M_{\rm BH}}\sqrt{M_{\rm BH}^2-\left(\frac{A_{\rm AH}}{8\pi M_{\rm BH}}-M_{\rm BH}\right)^2}
% %    a_{\rm{BH}}=\frac{1}{M_{\rm BH}}\sqrt{M_{\rm BH}^2-\left(\frac{A_{\rm AH}}{8\pi M_{\rm BH}}-M_{\rm BH}\right)^2}
%     \label{eq:chi_BH}
% \end{equation}
% \kenta{Kiuchi: Eqs.(8-9) are redundant because they are indential to Eqs.(6-7).}

From the comparison of different estimations of the spin parameter, it has been found that the systematic error with this method to estimate the spin is much less than 0.01~\cite{kyutoku2010aug,kyutoku2011sep,shibata2009feb}. Therefore, the spin parameter used in the paper for developing fitting formulas is accurate enough for our study. %\kenta{Kiuchi: we don't need "we believe".}

\subsubsection{Gravitational waves}

To derive gravitational waveforms, we extract the Weyl scalar $\Psi_4$ at the coordinate radius of $D=480 M_{\odot}$ from the coordinate origin and extrapolate them to the null infinity with a method from the BH perturbation theory~\cite{lousto2010nov}. The gravitational waveforms $h_{\rm GW}$ are obtained from the time integration of the $l=|m|=2$ spherical harmonics modes of $\Psi_4$. All the waveforms in this paper are shown for the observer along the $z$-axis. To reduce the nonphysical low-frequency components, we do not perform the time integration directly but adopt the method proposed by Reisswig and Pollney \cite{reisswig2011sep}.
%\kk{(!!!The presicription for interpolataing waveforms to the null ininity (Nakano method) should be also used!!!}
The retard time $t_{\rm ret}$ is approximately obtained by
\begin{equation}
    t_{\rm ret}=t-D-2m_0 \ln(D/m_0).
    \label{eq:t_ret}
\end{equation}
%where $m_0=M_{\rm{BH}}+M_{\rm{NS}}$ is the initial total gravitational mass of the system at infinite separation defined in \autoref{tab:initial_para}.

The GW spectrum is obtained as the sum of the Fourier components of the two polarizations of $l=|m|=2$ modes:
\begin{equation}
     \tilde{h}(f)=\sqrt{\frac{|\tilde{h}_+(f)|^2+|\tilde{h}_{\times}(f)|^2}{2}}, 
    \label{eq:hf1}
\end{equation}
where
\begin{equation}
     \tilde{h}_{+,\times}(f)=\int e^{2\pi ift}h_{+,\times}(t)\ dt. 
    \label{eq:hf2}
\end{equation}
We use the normalized amplitude $f|\tilde{h}(f)|D/m_0$ as a function of a dimensionless frequency $fm_0$ to show the GW spectrum in the following plots. 

\subsection{Fitting formulas}\label{sec:fitting}

Former studies give fitting formulas for $\hat{M}_{\rm{rem}}$~\cite{foucart2012dec,foucart2018oct} and $\hat{M}_{\rm{eje}}$~\cite{kawaguchi2016jun,krger2020feb}. In these fitting formulas, $\hat{M}_{\rm{rem}}$ and $\hat{M}_{\rm{eje}}$ are determined by the mass ratio $Q$, the dimensionless BH spin $\chi_{\rm{BH}}$, and the NS compactness $C_{\rm{NS}}$ except for the one derived in Ref.~\cite{kawaguchi2016jun}, which also depends on the value of $M_{\rm NS}/M_{\rm b}$. In Sec.~\ref{sec:fixed-C} we will demonstrate that this parametrization is sufficient for estimating $\hat{M}_{\rm{rem}}$ and $\hat{M}_{\rm{eje}}$. 

\begin{description}
\item[1] In Ref.~\cite{foucart2012dec}, the fitting formula for the remnant mass (referred to as rem\_2012) is determined as
\begin{align}\label{eq:rem_2012}
    \frac{M^{\rm{rem}}_{\rm{fit},2012}}{M_{\rm{b}}}&=\alpha(3Q)^{1/3}(1-2C_{\rm{NS}})-\beta\frac{R_{\rm{ISCO}}}{R_{\rm{NS}}}, \\
    &=\alpha(3Q)^{1/3}(1-2C_{\rm{NS}})-\beta r_{\rm{ISCO}}(\chi_{\rm{BH}})QC_{\rm{NS}},
\end{align}
where $\alpha$ and $\beta$ are determined by minimizing the $\chi$-square (as defined in Eq.~(\ref{eq:chi_square}) below) of the fitting formula, $R_{\rm{ISCO}}$ is the radius of the innermost stable circular orbit (ISCO) of the remnant BH with $r_\mathrm{ISCO}=R_\mathrm{ISCO}/M_\mathrm{BH}$, and other quantities are defined in Table~\ref{tab:quantity}.
The ranges of the initial NS and BH parameters covered by the numerical simulations are $Q=3$--$7$, $\chi_{\rm{BH}}=0$--$0.9$, $C_{\rm{NS}}=0.13$--$0.18$, and the best-fit model gives $\alpha\approx 0.288$ and $\beta \approx 0.148$. Here $R_{\rm{ISCO}}$ is given by
\begin{align}
    &Z_1=1+(1-\chi_{\rm{BH}})^{1/3}\nonumber \\
    &~~~~~~~~~~\times [(1+\chi_{\rm{BH}})^{1/3}+(1-\chi_{\rm{BH}})^{1/3}] \\
    &Z_2 = \sqrt{3\chi_{\rm{BH}}^2+Z_1^2} \\
    &\frac{R_{\rm{ISCO}}}{M_{\rm{BH}}}=3+Z_2\nonumber \\
    &~~~~~~~~~~~ -\rm{sign}(\chi_{\rm{BH}})\sqrt{(3-Z_1)(3+Z_1+2Z_2)},
\end{align}
as in Ref.~\cite{bardeen1972dec}.

\item[2]
Ref.~\cite{foucart2018oct} gives a modified fitting formula for the remnant mass (referred to as rem\_2018)
\begin{equation}
    \frac{M^{\rm{rem}}_{\rm{fit},2018}}{M_{\rm{b}}}= \\ \left[\rm{Max}\left(\alpha\frac{1-2C_{\rm{NS}}}{\eta^{1/3}}-\beta r_{\rm{ISCO}}\frac{C_{\rm{NS}}}{\eta}+\gamma, 0\right)\right]^{\delta},
\end{equation}
where $\eta$ denotes the symmetric mass ratio $\eta = Q/(1+Q)^2$. The best fitting gives $\alpha=0.406,\beta=0.139,\gamma=0.255,\delta=1.761$ and the parameter space for the simulations was $Q=1$--$7$, $\chi_{\rm{BH}}=-0.5$--$0.9$, $C_{\rm{NS}}=0.13$--$0.182$ and $M_{\rm{rem}}\leq 0.3M_{\rm{b}}$. %\kenta{Kiuchi: NS is not necessary in $M^{\rm{b}}$.}

\item[3]
Ref.~\cite{kawaguchi2016jun} gives the fitting formula for the ejecta mass (referred to as eje\_2016) as
\begin{align}
%\begin{split}
    \frac{M^{\rm{eje}}_{\rm{fit},2016}}{M_{\rm{b}}}&=\text{Max}\Big[a_1Q^{n_1}\frac{1-2C_{\rm{NS}}}{C_{\rm{NS}}}-a_2Q^{n_2}r_{\rm{ISCO}}(\chi_{\rm{eff}})\nonumber \\
    &~~~~~~~+a_3(1-\frac{M_{\rm{NS}}}{M_{\rm{b}}})+a_4,\ 0\Big] \label{eq:eje_2016}
%\end{split}
\\
    \chi_{\rm{eff}} &= \chi_{\rm{BH}}\cos{i_{\rm{tilt}}}
\end{align}
In our simulations, the angle between the BH spin and the orbital angular momentum is  $i_{\rm{tilt}}=0$. The best-fit gives $a_1=4.464\times10^{-2},\ a_2=2.269\times10^{-2},\ a_3=2.431,\ a_4=-0.4159,\ n_1=0.2497,\ n_2=1.352$, and the parameter space for simulations was $Q=3$--$7$, $\chi_{\rm{BH}}=0$--$0.75$, and $C_{\rm{NS}}=0.138$--$0.18$ by fixing $M_{\rm{NS}}=1.35M_{\odot}$ and changing EOS.

% What's more, Ref.~\cite{kawaguchi2016jun} also gives a fitting formula for the average velocity of the ejecta
% \begin{equation}
%     v_{\rm{ave}}/c=(0.01533Q+0.1907).
% \end{equation}
\item[4]
Ref.~\cite{krger2020feb} gives a modified fitting formula for the ejecta mass (referred to as eje\_2020)
\begin{align}
    \frac{M^{\rm{eje}}_{\rm{fit},2020}}{M_{\rm{b}}}=a_1Q^{n_1}\frac{1-2C_{\rm{NS}}}{C_{\rm{NS}}}-a_2Q^{n_2}r_{\rm{ISCO}}+a_4.
\end{align}
The fitting results are $a_1=0.007116,\ a_2=0.001436,\ a_4=-0.02762,\ n_1=0.8636,\ n_2=1.6840$, and the parameter space for simulations was the same as that of Ref.~\cite{kawaguchi2016jun}. 

\item[5]
Ref.~\cite{kawaguchi2016jun} also gives a fitting formula for the average ejecta velocity (cf, Eq.~(\ref{eq:v_eje})) as a simple linear function of $Q$:
\begin{equation}\label{eq:fit_veje}
    v^{\rm{eje}}_{\rm{fit,2016}}=(0.01533Q+0.1907)c.
\end{equation}
The definition of average ejecta velocity used to fit $ v^{\rm{eje}}_{\rm{fit,2016}}$ is given by Eq.~(\ref{eq:veje_old}) in Appendix~\ref{sec:appendix2}, which is not the same as $v_{\rm{eje}}$ given in Eq.~(\ref{eq:v_eje}).
\end{description}

\subsection{Models} \label{sec:models}
%\kk{(KK:more explanation for the reason and the purpose why these BH-NS configurations are studied needed to be explaned here.)}

\input{table_initial_para}

Table~\ref{tab:initial_para} lists BH-NS binary models and initial parameters used in our study. The models are labeled by EOS-$Q\chi_{\rm{BH}}M_{\rm{NS}}$, i.e. 15H-Q3a75M18 means the model with NS EOS ``15H", mass ratio $Q=3$, NS mass $M_{\rm{NS}}=1.8M_{\odot}$, and dimensionless BH spin $\chi_{\rm{BH}}=0.75$ aligned with the orbital angular momentum.
%and a grid resolution $N=82$.
%\kenta{Kiuchi: Is $N$ in the model name necessary? In many figures like Fig. 2 and 3, the information for $N$ is lacking.}

%Generally, tidal disruptions enhance as the spin goes up, and decrease as the mass ratio increases due to weaker tidal forces \cite{kyutoku2011sep}, therefore, among the three combinations, the $Q=3, a_{\rm{BH}}=0.75$ combination is supposed to have the strongest tidal disruption. 
%We consider circular binaries where the BH spin is parallel to the orbital momentum. 

In Sec.~\ref{sec:fixed-C}, to verify our hypothesis that the normalized disk mass and ejecta mass only depend on $Q$, $\chi_{\rm{BH}}$, and $C_{\rm{NS}}$, we perform simulations by varying the NS mass with fixing these three parameters, and explore whether the differences of results among different models are within the range of numerical accuracy. To compare our results with previous ones, we choose three configurations of $Q$ and $\chi_{\rm{BH}}$ as $(Q,\chi_{\rm{BH}})=\ (3,0.5),\ (3,0.75)$ and $(5,0.75)$. We study the models fixing the NS compactness with $C_{\rm{NS}}=0.182$, but employing EOS models with two-pieces 15H, HB, and four-pieces EOS models H4 and APR4~\cite{Akimal1998sep,glendenning1991oct,lackey2006jan} to vary the NS mass. Since the ejecta mass is small in the $C_\mathrm{NS}=0.182$ cases, we also perform simulations with fixed compactness $C_{\rm{NS}}=0.147$, with two-pieces EOS models 15H, H, B, and four-pieces EOS model APR4, for which we expect to get a larger disk and ejecta masses. In the following, we will refer to the models with fixed compactness $C_{\rm{NS}}=0.147$ and $C_{\rm{NS}}=0.182$ as C147 and C182, respectively.
%Note that this is already investigated for the cases with the NS mass of $\approx1.4\,M_\odot$ \cite{kyutoku2011sep,kyutoku2015aug} 
% \kk{(Please add appropriate references here, if avilable)}, but not for larger NS mass cases, in which the high density part of the EOS plays a role\wlh{(this consistency of two piecewise and four piecewise EOS for M less than 1.4Msun can be seen directly from the definition of EOS. I have added an analyze for the result in Sec.~\ref{sec:fixed-C_mass}, so I think this sentence can be deleted.)}\kk{(I agree with this comment for low NS mass cases. Modified the sentence to mention that consistency is not investigated for large values of NS masses.)}. 

%\kk{(!!!the purpose for studying the models with 4-piece-wise EOS should be mentioned here.!!!)}
%Results in Sec.~\ref{sec:fixed-C} show that given the same $Q$ and $a_{\rm{BH}}$, the remnant and the ejecta normalized by the baryon mass of the NS are dominated by $C_{\rm{NS}}$, irrespective of the Arnowitt-Deser-Misner (ADM) mass of the NSs. As a result, it is reasonable for us to study the tidal disruption for more compact NSs by fixing $M_{\rm {NS}}=1.8M_{\odot}$ in Sec.~\ref{sec:mNS}. 
We note that the purpose of studying the models with four-pieces EOS is to test whether the simplified two-pieces EOSs can be used to accurately describe the NS tidal disruption in BH-NS mergers and to test how the difference in the detailed internal structure of the NS can change the results of the simulations. Two-pieces polytopes are simplified EOSs, which are typically employed in addressing less massive NSs. It can be proved that for an NS with mass less than $1.4M_{\odot}$, its central density is relatively low, which means the EOS at higher density plays a minor role. Hence, two-pieces and four-pieces EOS should provide approximately the same results for the C147 models. On the other hand, for the C182 models, the NS structure in high-density regions can be sensitive to the values of $\Gamma_2$ and $\Gamma_3$. Therefore, it is necessary to perform simulations with both two-pieces and four-pieces EOS models to test whether our conclusions can be applied despite the detailed structure of the NS.

In Sec.~\ref{sec:mNS}, we further perform simulations with a high value of $M_{\rm{NS}}=1.8M_{\odot}$ while fixing $Q$ and $\chi_\mathrm{BH}$ to check the validity of the fitting formulas with large values of NS compactness. 
%The corresponding $M_{\rm{BH}}$ for $Q=3$ and $5$ in this NS mass are $5.4M_{\odot}$ and $9.0M_{\odot}$, respectively. 
As we demonstrate in Sec.~\ref{sec:fixed-C} that merger dynamics depends on $M_{\rm{NS}}$ and $R_{\rm{NS}}$ approximately only through the NS compactness ($C_{\rm NS}=M_{\rm{NS}}/R_{\rm{NS}}$), it is an interesting exploration to see the dependence of the merger dynamics by varying the compactness of NSs while fixing $M_{\rm{NS}}$, i.e. by changing the EOS of NSs. We choose EOSs, 15H, H, and B, with the compactness varying from 0.194 to 0.252. 

To evaluate the validity of fitting formulas, we calculate the $\chi$-square of their fittings, with the estimation of numerical error from both Refs.~\cite{kawaguchi2016jun,foucart2012dec} and our study. Numerical simulations are performed with grid resolutions $N=62,\ 82$ for all the models, and $N=102$ for some of them (see Appendix~\ref{sec:appendix1}) to estimate the numerical errors due to the finite grid resolution. We take $N=82$ as the fiducial resolution of our simulations and other results are summarized in the Appendix~\ref{sec:appendix1}.

% We also perform simulations for the same model with different grid resolutions N62 and N82. For some models, we perform an additional N102 run. We take N82 runs as our fiducial runs in this paper. The physical quantities in tables and figures, unless specified, are taken from models with a grid resolution $N=82$. The analysis of N62 and N102 runs can be seen in the Appendix~\ref{sec:appendix1}.

\section{Results} \label{sec:results}

\begin{figure*}[t]
\centering
\includegraphics[width=\columnwidth]{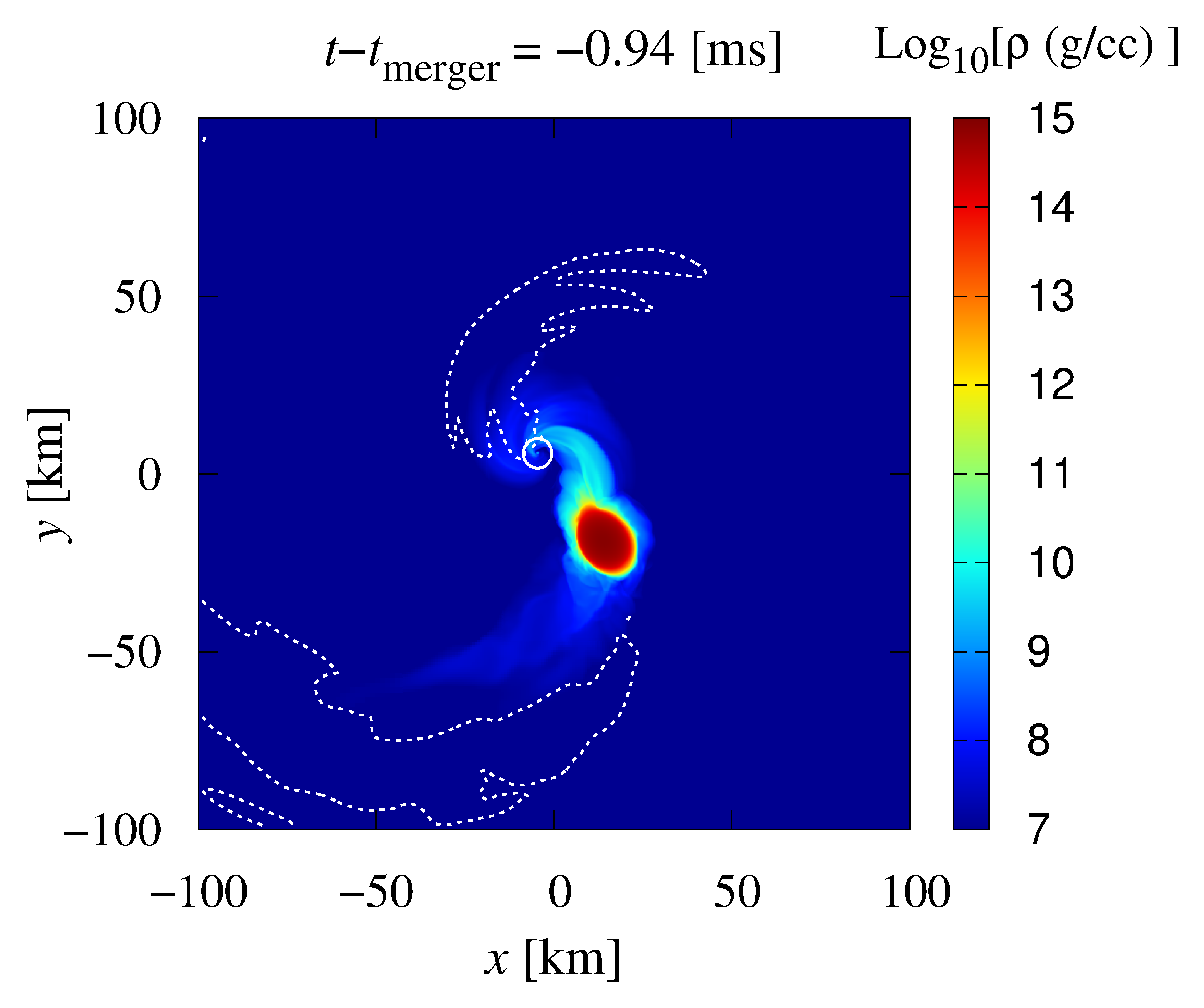}
\includegraphics[width=\columnwidth]{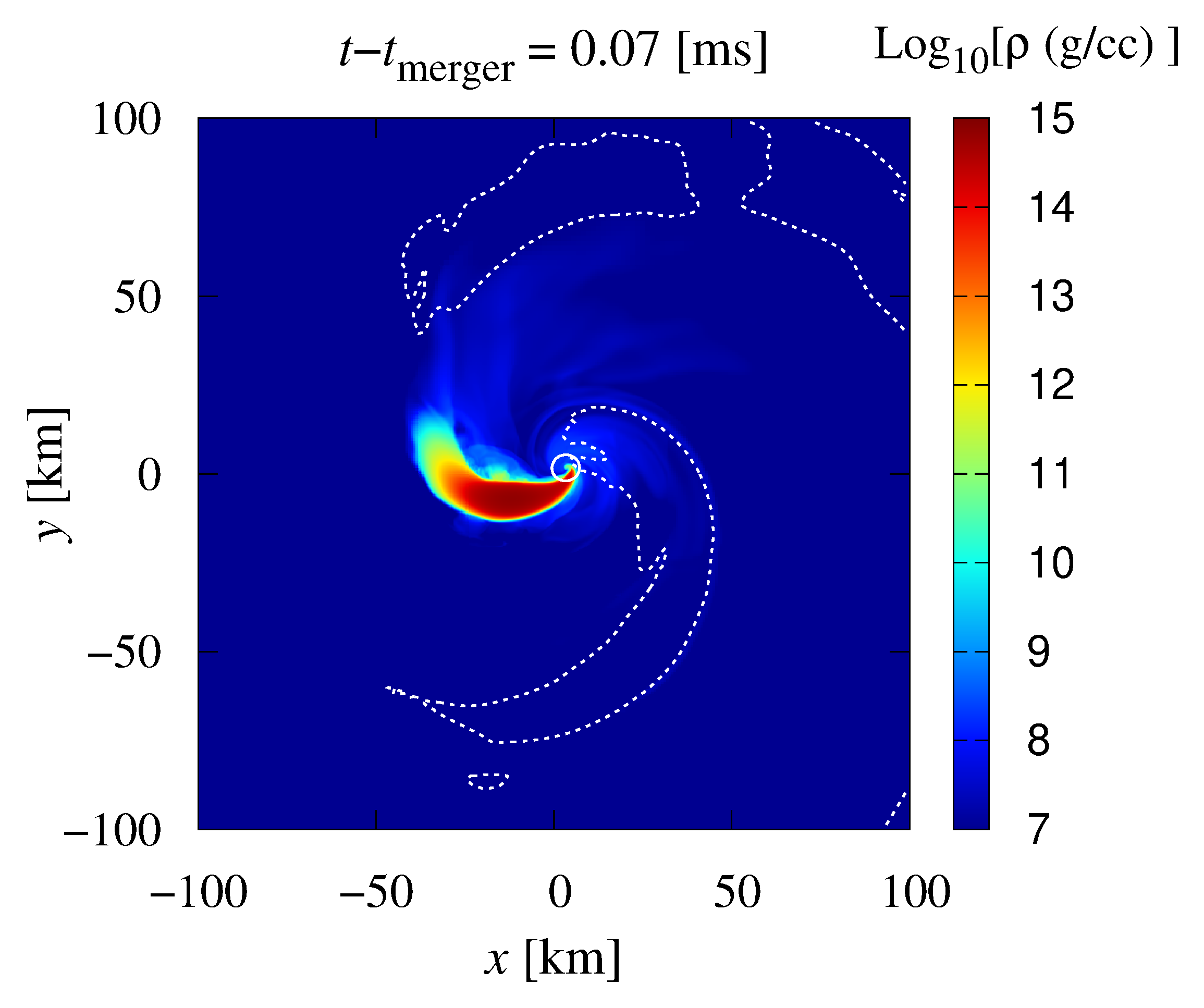}\\
\includegraphics[width=\columnwidth]{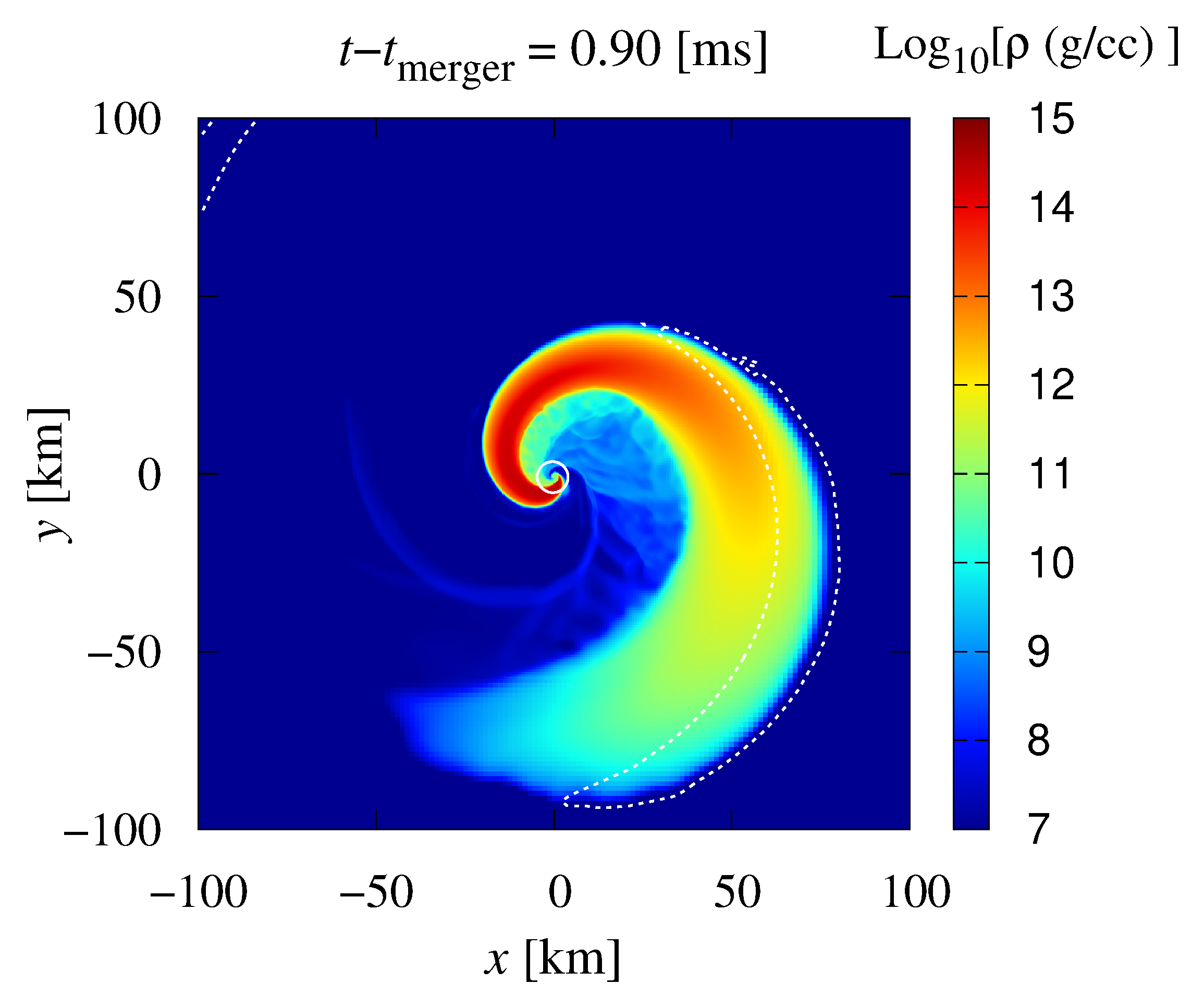}
\includegraphics[width=\columnwidth]{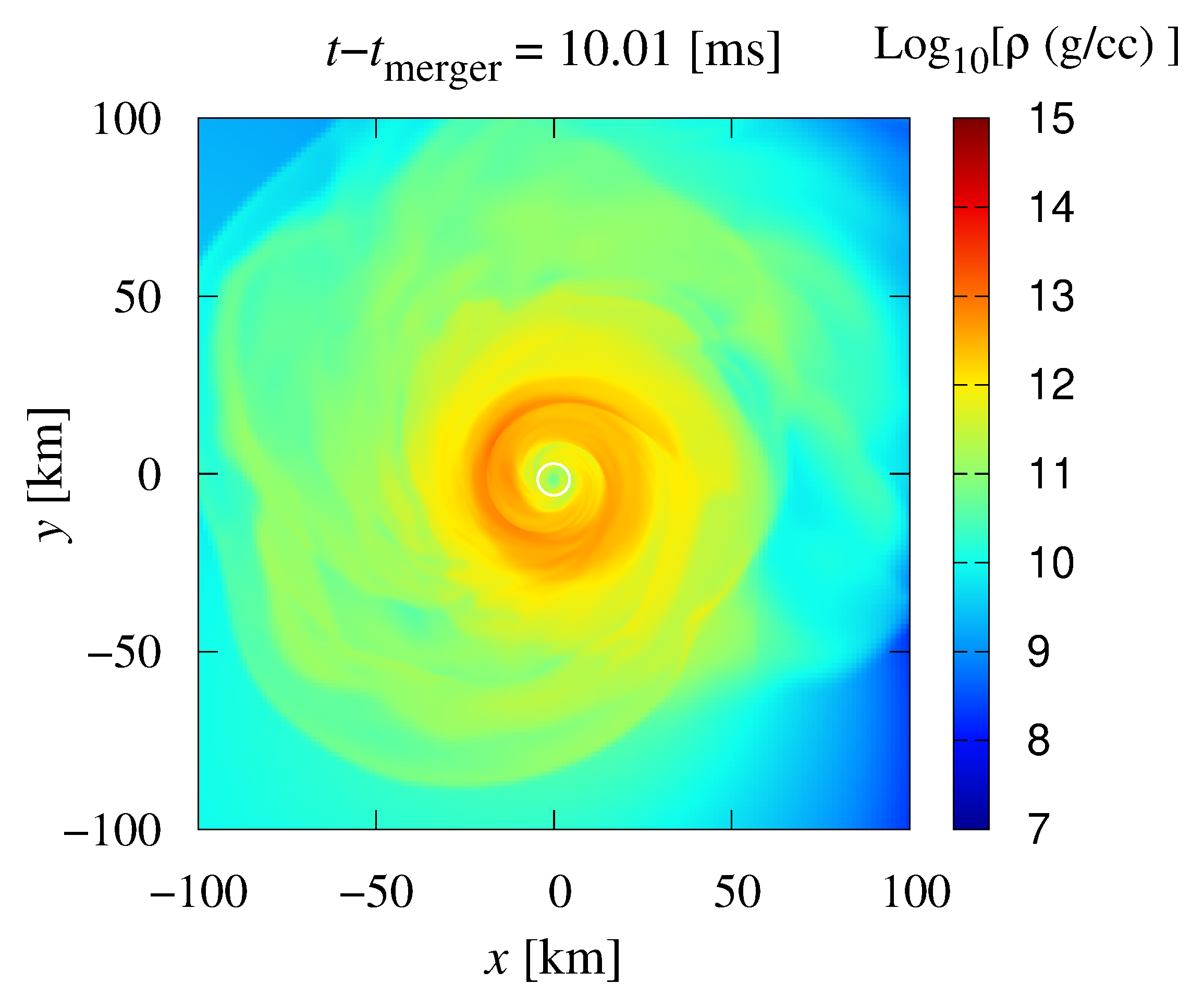}\\
\caption{Time evolution of the rest-mass density profile for model HB-Q3a75M1428 with $N=82$ run at $t-t_{\rm merger} \approx -0.94\,\rm{ms}$ (top left), $0.07\,\rm{ms}$ (top right), $0.90\,\rm ms$ (bottom left), and $10.01\ \rm ms$ (bottom right). The white circle indicates the apparent horizon. The white dashed lines show the region of unbound components in which the relation $-u_t\geq1$ is satisfied. }
\label{fig:dens_profile}
\end{figure*}

We present our numerical results in this section, focusing primarily on their dependence on the NS compactness, the mass ratio, and the dimensionless BH spin. After we describe an overview of the BH-NS merger process in Sec.~\ref{sec:overview}, we summarize the result for the case of a fixed NS compactness in Sec.~\ref{sec:fixed-C}, and then, for the case of massive and compact NSs in Sec.~\ref{sec:mNS}. The numerical results are shown in Tables~\ref{tab:C18}, \ref{tab:C147}, and \ref{tab:M18}, respectively. %The physical quantities in all the tables are measured at 10 ms after the merger.

\subsection{Overview of the merger} \label{sec:overview}

First we briefly describe an overview of the BH-NS merger process in our numerical simulations. Broadly speaking, we find the same features as for $M_{\rm NS}\approx 1.4\,M_\odot$ studied in the previous works \cite{kyutoku2010aug,kyutoku2011sep,kyutoku2015aug,shibata2006dec,shibata2007may,shibata2008apr,shibata2009feb,foucart2019may,hinderer2019sep,etienne2008apr,duez2008nov,etienne2009feb,chawla2010sep,duez2010may,foucart2011jan,foucart2012feb,foucart2012dec,etienne2012mar,etienne2012oct,foucart2013apr,lovelace2013jun,deaton2013sep,foucart2014jul,paschalidis2015jun,kawaguchi2015jul,kawaguchi2016jun,foucart2017jan,kyutoku2018jan,brege2018sep,ruiz2018dec,foucart2019feb}: 
The BH-NS mergers can have two types of fate. One is that the NS is tidally disrupted by the BH, and subsequently, an accretion disk is formed around the remnant BH and a part of the NS matter is ejected as the dynamical ejecta; the other is that the NS simply plunges into the BH with no tidal disruption and no mass ejection. In the following, we pay attention only to the former case because it can provide a lot of information about the BH-NS systems through the cutoff frequency and EM counterparts powered by the ejecta. 

The condition of tidal disruption can be obtained by comparing the ISCO radius $R_{\rm ISCO}$ and the mass-shedding radius $R_{\rm dis}$. If $R_{\rm ISCO}\leq R_{\rm dis}$, the NS can be tidally disrupted before it is swallowed into the BH. Assuming Newton's gravity for simplicity, we have \cite{shibata2011aug}
\begin{equation} \label{eq:R_dis_minus_R_isco}
    \frac{R_{\rm{dis}}}{R_{\rm ISCO}}\propto\frac{1}{C_{\rm{NS}}Q^{2/3}r_{\rm{ISCO}}(\chi_{\rm{BH}})}.
\end{equation}
This ratio approximately separates the two types of final fates mentioned above. If the ratio is larger than unity, the NS is expected to be tidally disrupted by the BH. Otherwise, the BH-NS binaries encounter an ISCO before reaching mass shedding limit, NS will simply plunge into the BH with no tidal disruption. Thus, the significance of the tidal disruption appears to depend on $Q$, $\chi_\mathrm{BH}$, and $C_{\rm{NS}}$, and the tidal disruption is less relevant for larger values of $Q$ and $C_{\rm{NS}}$ and smaller value of $\chi_{\rm{BH}}$ since $R_{\rm{ISCO}}$ decreases as $\chi_{\rm{BH}}$ increases. However, the relation of this ratio may be modified in general relativity, in particular for high-mass NSs for which the general relativistic effect is enhanced. The dependence of ejecta mass and the remnant mass on these three parameters as well as on the NS mass are discussed in detail in Secs.~\ref{sec:fixed-C} and \ref{sec:mNS}. 

Figure~\ref{fig:dens_profile} shows the snapshot of the rest-mass density profile, unbound matter, and the apparent horizon at selected time slices for model HB-Q5a75M1428. The separation between the NS and BH decreases due to GW emission, and after it reaches the tidal disruption radius, the NS is tidally disrupted (see the top panels in Fig.~\ref{fig:dens_profile}). While most of the disrupted NS matter is swallowed by the BH, a fraction of the matter will gain angular momentum by gravitational torque to form the tidal tail which remains outside the BH. The gravitationally-bounded matter in the tidal tail gradually falls back and forms a remnant disk around the remnant BH (see the bottom panels in Fig.~\ref{fig:dens_profile}), while the matter in the outer part of the tidal tail becomes dynamical ejecta. 

\begin{figure}[t]
	\includegraphics[width=0.45\textwidth]{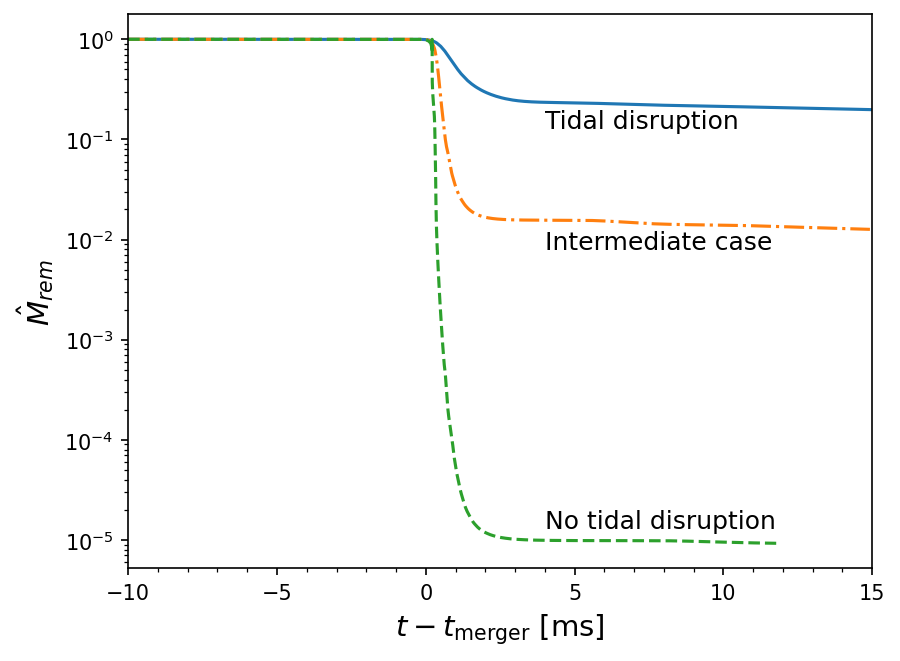}
    \caption{Time evolution of $\hat{M}_{\rm{rem}}$ for representative models, H-Q3a75M1220, 15H-Q3a5M18, and H-Q5a75M18 (solid, dot-dashed, and dashed lines,  respectively). Each model represents the cases with significant tidal disruption, the intermediate situation, and no tidal disruption, respectively. %Two vertical dotted lines show the onset time of the merger and at 10\,ms after the merger,  respectively.
    %\kenta{Kiuchi: The x-axis label should be $t-t_{\rm merger}$, and the two vertical dotted lines are redundant. Instead of the legend for it, the legend for the model is useful. Please enlarge the font.}
    }
        \label{fig:rem-t}
\end{figure}

Figure~\ref{fig:rem-t} shows the time evolution for the rest mass of the matter located outside the BH for three models, H-Q3a75M1220, 15H-Q3a5M18, and H-Q5a75M18 models, which represent the cases with significant tidal disruption, the intermediate case, and no tidal disruption, respectively. It shows that in tidal disruptions, indeed most of the NS matter is swallowed by the BH soon after the onset of the merger, i.e., within $\sim$1\,ms, and the rest mass of the matter remaining outside the BH depends strongly on the binary parameter. For the case with significant tidal disruption (H-Q3a75M1220), 
1\%--20\% %\kenta{Kiuch: Did you mention about 15H-Q3a5M18 case as well? Otherwise, it should be $\approx$20\% from Fig. 2.}
of the mass surrounds the BH to form a remnant disk, but only less than $0.01\%$ of matter remains outside the BH after the merger for the no tidal disruption case (H-Q5a75M18).

\begin{figure}[t]
	\includegraphics[width=0.45\textwidth]{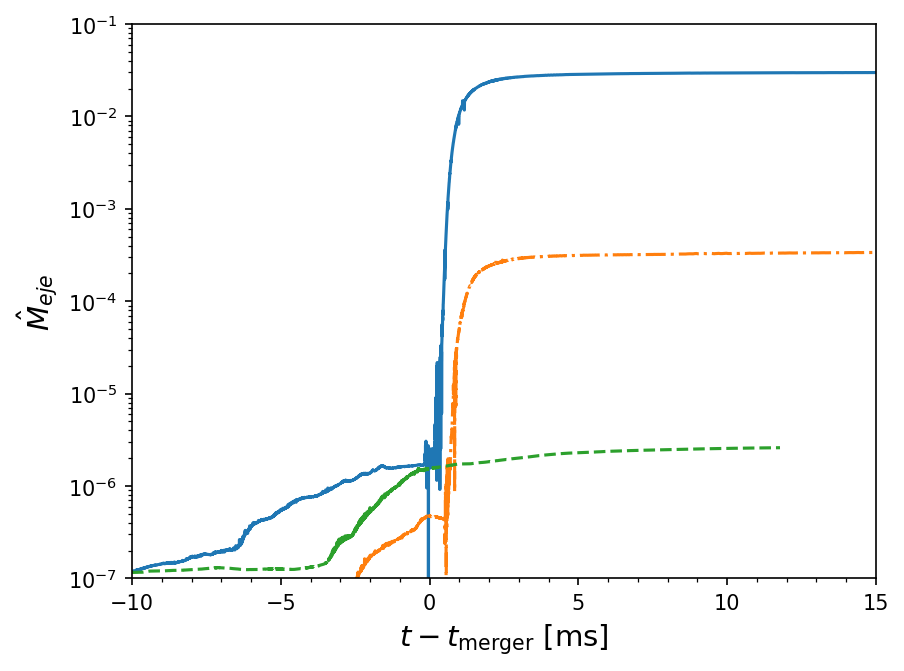}
    \caption{Time evolution of $\hat{M}_{\rm{eje}}$ for the same models as in Fig.~\ref{fig:rem-t}. 
    The $y$-axis shows the normalized ejecta mass. The tiny amount of ejecta before and right after the merger is due to the mass conservation error around the NS atomsphere.%\kk{(!!!Please remove the arrows with the label "ejecta before merger" as it is not the focus of our explanation and study. Also, the 15HQ3a5M18 model is not a good representative for discussing ejecta: It will be better to show the same models as in Fig.~\ref{fig:rem-t}. !!!)}
    %\kenta{Kiuchi: The x-axis label should be $t-t_{\rm merger}$. The two vertical dashed lines and the last sentence are redundant. Also, please mention about the baryon mass conservation error. Otherwise, the reader may be unable to judge whether the tiny amount of the ejecta in the orange or green curves has meaning. Please enlarge the font.}
    }
    \label{fig:eje-t}
\end{figure}

At the time of the merger, a fraction of the matter outside the BH becomes gravitationally unbound, i.e., the dynamical ejecta. The region in Fig.~\ref{fig:dens_profile} surrounded by the dashed curves denotes the matter with gravitationally unbound orbits. Figure~\ref{fig:dens_profile} shows that the ejecta is formed around the outer edge of the tidal tail. The time evolution of the ejecta mass is shown in Fig.~\ref{fig:eje-t}. The ejecta are formed within the dynamical time scale ($\sim 1$\,ms) after the onset of the merger, and, as is the case for the remnant matter located outside the BH, the amount of the ejecta depends strongly on the binary parameters.

%The material outside the apparent horizon can be classified as bounded and unbounded materials, i.e. ejecta.
%Mergers with tidal disruptions can launch relativistic ejecta, which are neutron-rich and can produce strong X-ray and radio emissions. The mechanism of relativistic jets has not been fully figured out yet, and the jets possibly extract the rotation energy of the BH from the ergosphere of a spinning BH through its magnetic field and become unbounded (reference). 
%The time evolution of the ejecta mass is shown in Fig.~\ref{fig:eje-t}. Unlike $M_{\rm{rem}}$, which increases immediately after the onset of the merger, there is a bulk of ejecta before the time of the merger, due to numerical dissipation around the atmosphere of the NSs. This contributes a lot to the uncertainties of the ejecta mass, and changes as the Riemann solver used in the simulation changes (see Appendix~\ref{sec:appendix1}) \wlh{(where is this part exactly?)}.
%(cite: (Blandford and Znajek 1977).) 

%\kk{(KK: This section is for explaining the overview. It is better to not mixed up the sentence to explain the purpose of study here. )}

%\kk{(KK: It will be good have the explanation for the evolution of remnant mass and ejecta. For example, readers may have the question why the mass outside the AH keeps decreasing after the merger. )}
%\input{table2}

\subsection{Fixed compactness case} \label{sec:fixed-C}

%\kk{(Tables are quite messy. We can split the table into those which explains the initial set up of the models and the results for the remnant. The way of presenting the results with different resolution can be recinsidered to slim up the table: For example, you can present the results of ejecta mass for diferent resolutions with the style XX(N62)/XX(N82)/XX(N102) in the same cell with explanation on the header. )}

\subsubsection{Properties of remnant disks and ejecta} \label{sec:fixed-C_mass}

% To verify our hypothesis that the disk mass and ejecta mass, normalized by the NS baryon mass, only depend on $C_{rm{NS}}, a_{\rm{BH}}$ and $Q$. We perform simulations by fixing these three parameters and changing the EOS, i.e. changing the NS mass. We first fix the compactness to be $C_{\rm{NS}}=0.182$, and choose other parameters to be $Q=3, a=0.5$, $Q=3, a=0.75$, $Q=5, a=0.75$.

 %\sout{Our simulation results differ from theirs a lot. \wlh{(why??)} Conclusions in this paper are derived only from our simulations}
% , with two piecewise EOS models ``15H, HB" and four piecewise EOS models ``H4, APR4". We use $\hat{M}$ to denote $M/M_{\rm{b}}$. Due to the small ejecta mass in C182 cases, we further perform simulations with fixed compactness $C_{\rm{NS}}=0.147$, with EOS ``H, B, APR4, H4", which have larger disks and ejecta mass. 
\input{table_result_C18}
\input{table_result_C147}

We list the key results of the simulations for C182 and C147 models %\kenta{Kiuchi: Since you already define C182 and C147, you don't have to repeat like $C_{\rm NS}=0.182$ model.}
in Tables~\ref{tab:C18} and~\ref{tab:C147}, respectively. We also include the results for $\hat{M}_{\rm{rem}}$ and $\hat{M}_{\rm{eje}}$ with the same compactness from  Refs.~\cite{kyutoku2015aug,kyutoku2011sep}. Tables~\ref{tab:C18} and~\ref{tab:C147} show that $\hat{M}_{\rm{rem}}$ increases as $\chi_{\rm BH}$ increases, and as the compactness $C_{\rm{NS}}$ and the mass ratio $Q$ decreases. This is consistent with the predictions given in Sec.~\ref{sec:overview}. % and the results of the previous studies for the NS mass of $\approx1.4\,M_\odot$~\cite{kyutoku2015aug,kyutoku2011sep} \kk{(Please add more references.)}.
%This is reasonable since higher $a_{\rm BH}$ induces a smaller innermost stable circular orbit, and smaller compactness means less bounded by self-gravity, which all tend to strengthen the tidal disruption. $\hat{M}_{\rm{rem}}$ increases as mass ratio $Q$ decreases in the case of $C_{\rm{NS}} \approx 0.182$, which is reasonable since higher $Q$ induces weaker tidal disruption as shown in the analysis in Sec.~\ref{sec:overview}. 
However, we note that the dependence of $\hat{M}_{\rm{rem}}$ on $Q$ is rather weak in the case of the lower compactness $C_{\rm{NS}} \approx 0.147$, which deviates from the naive prediction of Newton approximation shown in Eq.~(\ref{eq:R_dis_minus_R_isco}). This weaker dependence is also found and an interpretation for this is described in Ref.~\cite{hayashi2021feb}. 

%before but has not been fully understood yet~\cite{hayashi2021feb}.

From Table~\ref{tab:C147} and \ref{tab:C18}, it is found that $\hat{M}_{\rm{eje}}$ increases as $\chi_{\rm BH}$ increases and as $C_{\rm{NS}}$ decreases. This is the same dependence as $\hat{M}_{\rm{rem}}$. However, being different from $\hat{M}_{\rm{rem}}$, $\hat{M}_{\rm{eje}}$ increases as $Q$ increases for the lower compactness $C_{\rm{NS}} \approx 0.147$. This has also been found in the previous works~\cite{shibata2009feb,kawaguchi2015jul,kyutoku2015aug,foucart2019may,foucart2012dec,foucart2014jul,kawaguchi2016jun}. This implies that the ejecta mass depends more strongly on the dynamical process of the tidal disruption than the remnant mass, and indicates that the dynamics of the tidally disrupted matter in the vicinity of the BH plays a key role in determining the ejecta mass.

%\kk{(Please add more references.)}
%This is probably because higher $Q$ induces more violent tidal disruption. \wlh{(why?)}

\begin{figure}[t]
    \centering
    \includegraphics[width=0.45\textwidth]{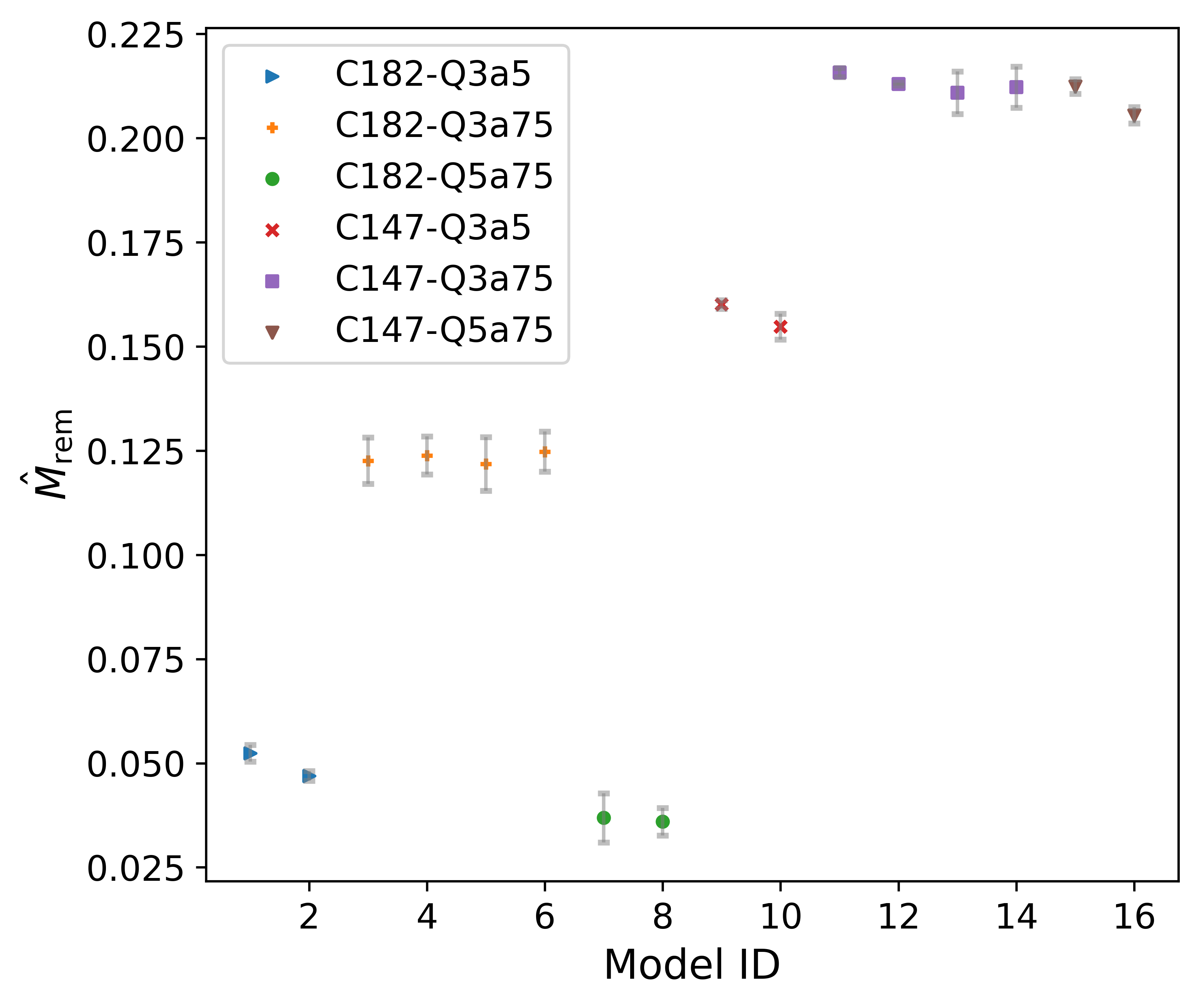}
    \caption{The values of $\hat{M}_{\rm{rem}}$ for the models with $C_{\rm{NS}}=0.182$ and $0.147$. The models with the same values of $\chi_{\rm{BH}}, Q$, and $C_{\rm{NS}}$ but with different NS mass $M_{\rm{NS}}$ are grouped with the same colors. The error bar of each model is determined by the difference between $N=62$ and $N=82$ results. Model ID corresponds to the column model ID in Tables~\ref{tab:C18} and \ref{tab:C147}. %\kenta{Kiuchi: "the error bar" in the first sentence is redundant. Please enlarge the font.}
    }
    \label{fig:fixedC_rem}
\end{figure}

Fig.~\ref{fig:fixedC_rem} compares the values of $\hat{M}_{\rm{rem}}$ among the models with the same values of $Q$, $\chi_{\rm{BH}}$, and $C_{\rm{NS}}$ but with different NS mass $M_{\rm{NS}}$, i.e., with different NS EOSs. The error bars are taken to be the discrepancy of the results between $N=62$ and $N=82$ runs of the same model. 
Hydrodynamic quantities do not always show monotonic dependency on the gird resolution and it is difficult to evaluate the error by checking the convergence.
Therefore we take the error estimation method shown above.
We note that the discrepancies between $N=82$ and $N=102$ results are mostly smaller than those between $N=62$ and $N=82$ results (see Appendix.~\ref{sec:appendix1}).
%\kenta{Kiuchi: Question. How reliable is the error estimation with this method? Normally, we should check the convergence with $N=62$, 82, and 102. In other words, if you plot $|M_{82}-M_{102}|/M_b$ on top of the upper panels in Fig. 12, how good is the error estimation given by the dashed curves?} 
%\wlh{Luohan: $|M_{82}-M_{102}|$ is generally smaller or equal to $|M_{82}-M_{62}|$, as shown in Table.~\ref{tab:convergence} and Fig.~12.}
If the range of error bars for different models with the same values of $\chi_{\rm{BH}}, Q$, and $C_{\rm{NS}}$ overlaps, we consider that the results are in agreement within the range of the accuracy. 

It is shown that, within the numerical accuracy, $\hat{M}_{\rm{rem}}$ with the same values of $Q$, $\chi_{\rm{BH}}$, and $C_{\rm{NS}}$ agrees with each other regardless of their different NS masses. %\wlh{This consistency is in accordance with our expectations, because as a quantity determined by the strength of tidal disruptions, $\hat{M}_{\rm{rem}}$ is expected to be dominated by the ratio of $R_{\rm{dis}}$ and $R_{\rm{ISCO}}$. }
We also find from Tables~\ref{tab:C18} and \ref{tab:C147} that the dimensionless spin of the remnant BH, $\chi_{\rm{BH}}^f$, is approximately the same among the models with the same initial values of $Q$, $\chi_{\rm{BH}}$, and $C_{\rm{NS}}$. This indicates that the BH swallows approximately the same amount of the mass and angular momentum normalized by $M_{\rm{NS}}$. These results show that $\hat{M}_{\rm{rem}}$ and the radius at which the NS tidal disruption occurs depends sensitively on $Q$, $\chi_{\rm BH}$, and $C_{\rm{NS}}$ while $M_{\rm{NS}}$ plays a role only in scaling the system.

\begin{figure}[t]
    \centering
    \includegraphics[width=0.45\textwidth]{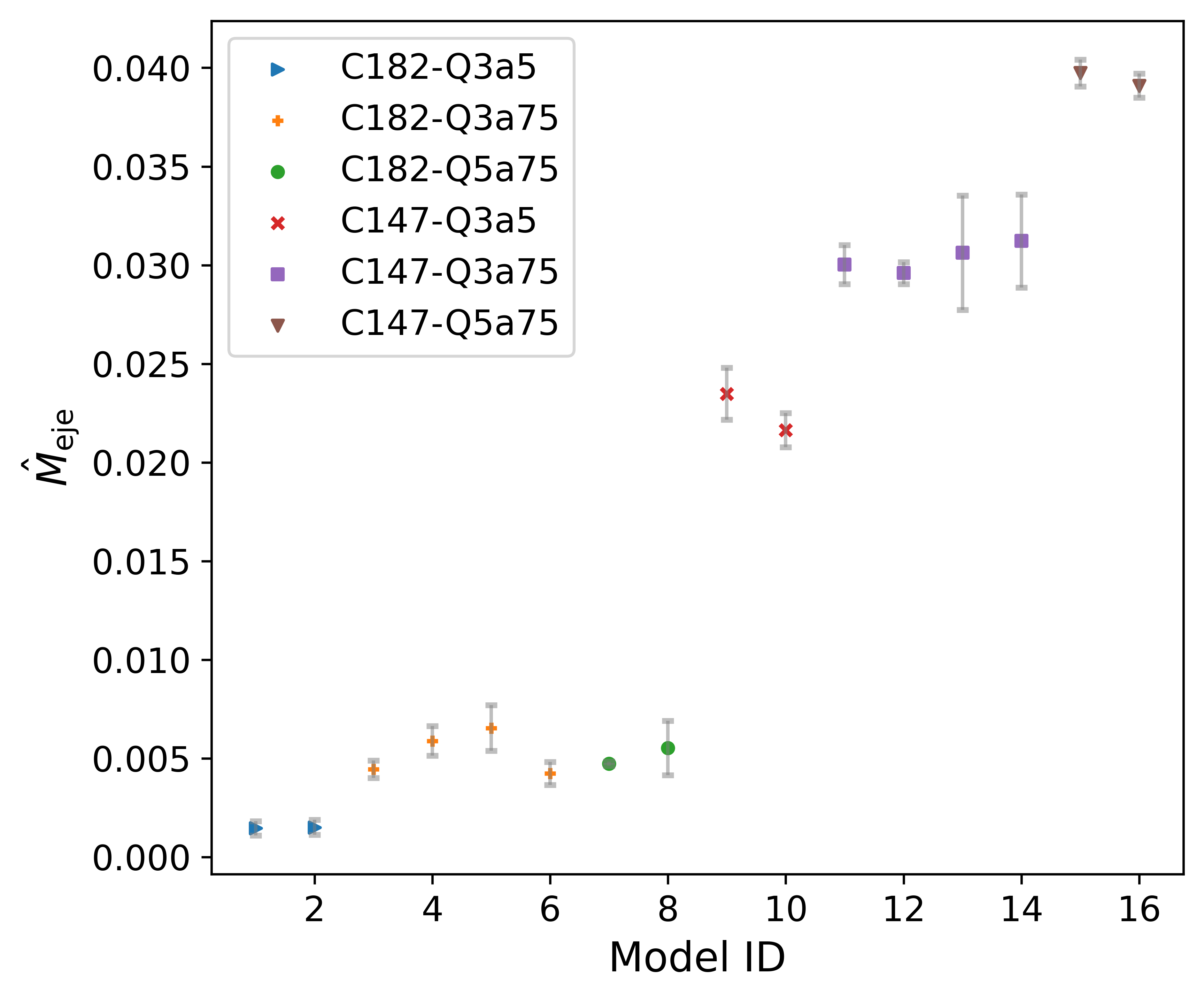}
    \caption{The values of $\hat{M}_{\rm{eje}}$ for models with $C_{\rm{NS}}=0.182$ and $0.147$. The models with the same values of $\chi_{\rm{BH}}, Q$, and $C_{\rm{NS}}$ but with different NS mass $M_{\rm{NS}}$ are grouped with the same colors. The error bar of each model is determined by the difference between $N=62$ and $N=82$ results. Model ID corresponds to the column model ID in Tables~\ref{tab:C18} and \ref{tab:C147}. %\kenta{Kiuchi: "the error bar" in the first sentence is redundant. Please enlarge the font.}
    }
    \label{fig:fixedC_eje}
\end{figure}

Fig.~\ref{fig:fixedC_eje} shows the results for the ejecta mass, with its error bar defined in the same way as Fig.~\ref{fig:fixedC_rem}. 
%It can be seen that aside from C182-Q3a75 models, which have deviations a little larger than the range of the error bar, other models all have results within the range of the error bars. Hence, while it is slightly more obscured by the numerical error than the case of $\hat{M}_{\rm{rem}}$, 
$\hat{M}_{\rm{eje}}$ with the same values of $Q$, $\chi_{\rm{BH}}$, and $C_{\rm{NS}}$ also agrees with each other. The results show that $\hat{M}_{\rm{eje}}$ also depends sensitively on $Q$, $\chi_{\rm BH}$, and $C_{\rm{NS}}$ but only minorly on $M_{\rm{NS}}$.
% Aside from $\hat{M}_{\rm{rem}}$ and $\hat{M}_{\rm{eje}}$, if we look at other quantities in Tables.~\ref{tab:C18},\ref{tab:C147}, 

%Fig.~\ref{fig:fixedC_eje} shows the results the same as Fig.~\ref{fig:fixedC_rem} but for the ejecta mass. It can be seen that aside from C182-Q3-a75 models having deviations a little larger than the range of the error bar, other models all have results within the range of error. %Therefore, though not as good as $\hat{M}_{\rm{rem}}$, the behavior of $\hat{M}_{\rm{eje}}$ also follows the same rule approximately, and our method of fixing $M_{\rm{NS}}$ and changing EOS also holds for the ejecta.

%\kk{(!!!Please explain what we can find from the results of the models with 4-piece-wise EOS and discuss the implication of them here.!!!)}

It is worth noting that the values of $\hat{M}_{\rm{rem}}$ and $\hat{M}_{\rm{eje}}$ for the models which employ the four-pieces EOSs (model ID $=5, 6$, and 14) are also approximately in agreement with the results for the models which employ the two-pieces EOSs with the same values of $Q$, $\chi_{\rm BH}$, and $C_{\rm{NS}}$. This implies that the detailed structure, i.e., the value of $\Gamma_2$ in Table~\ref{tab:EOS}, has only a minor effect on the results of $\hat{M}_{\rm{rem}}$ and $\hat{M}_{\rm{eje}}$. This conclusion also justifies us to study NSs with $M_\mathrm{NS}=1.8M_{\odot}$ using two-pieces polytropes in Sec.~\ref{sec:mNS}.

The fitting formula for the ejecta mass, Eq.~(\ref{eq:eje_2016}), given in Ref.~\cite{kawaguchi2016jun} includes not only $C_{\rm{NS}}$, $Q$, and $\chi_{\rm BH}$, but also the specific binding energy. Though the binding energy may physically have influence on the result, it is still consistent with our results. Since the specific binding energy can be well approximated as a function of the NS compactness $C_{\rm{NS}}$ (e.g., Ref.~\cite{lattimer2001mar}), Eq.~(\ref{eq:eje_2016}) can be effectively approximated as a function of only $C_{\rm{NS}}$, $Q$, and $\chi_{\rm BH}$. Also, we find that the variation of the specific binding energy term in our parameter space is nevertheless less than the estimated errors in our simulations (around 10\%).

%These results show that, as is found in the previous studies\kk{(Please add appropriate references here if available)}, the two piece-wise EOSs can be a good approximation for describing the physical process of NS tidal disruption in BH-NS mergers within the numerical errors of our simulations, especially for less massive NSs, and the detailed structure of the NS has only a minor effect on the values of $\hat{M}_{\rm{rem}}$ and $\hat{M}_{\rm{eje}}$ for the wide range of the NS mass.

\begin{figure}[t]
    \centering
    \includegraphics[width=0.45\textwidth]{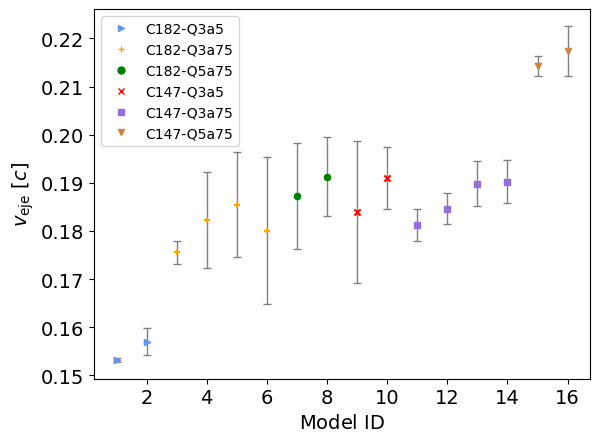}
    \caption{The value of $v_{\rm{eje}}$ at 10\,ms after the merger for models with $C_{\rm{NS}}=0.182$ and $0.147$. The models with the same values of $\chi_{\rm{BH}}, Q$, and $C_{\rm{NS}}$ but with different NS mass $M_{\rm{NS}}$ are grouped with the same colors. The error bar of each model is determined by the difference between $N=62$ and $N=82$ results. Model ID corresponds to the column model ID in Tables~\ref{tab:C18} and \ref{tab:C147}.}
    %\kenta{Kiuchi: "the error bar" in the first sentence is redundant. Please enlarge the font.}
    \label{fig:fixedC_veje}
\end{figure}

Fig.~\ref{fig:fixedC_veje} shows the average ejecta velocity $v_{\rm eje}$ and their error bars. Broadly speaking, $v_{\rm{eje}}$ increases with $Q$ and depends weakly on other parameters. This is consistent with the results of the previous studies for $M_\mathrm{NS} \approx1.4\,M_\odot$~\cite{kawaguchi2016jun,kyutoku2015aug,foucart2019may,foucart2017jan},
%(\kk{Please add appropriate references})
which is also described in fitting formula of Eq.~(\ref{eq:fit_veje}). As is the case for $\hat{M}_{\rm{rem}}$ and $\hat{M}_{\rm{eje}}$, given the same values of $Q$, $\chi_{\rm{BH}}$, and $C_{\rm{NS}}$, the values of $v_{\rm{eje}}$ for different models approximately agree with each other within the numerical accuracy.

The coefficients of the fitting formula of Eq.~(\ref{eq:fit_veje}) given in Ref.~\cite{kawaguchi2016jun} were obtained by fitting the average ejecta velocity $v_{\rm eje,old}$ obtained by various configurations of BH-NS binaries. They assumed a linear relation between $v_{\rm eje,old}$ and $Q$: $v_{\rm eje,old}=(a_1Q + a_2)c$, and used the least squares fitting to derive the coefficients $a_1$ and $a_2$. However, the definition of $v_{\rm eje,old}$ in Ref.~\cite{kawaguchi2016jun}, which is different from the present definition in Eq.~(\ref{eq:v_eje}), did not take into account the gravitational potential. Thus, the value of $v_{\rm eje,old}$ can be overestimated. %\kenta{Kiuchi: Why don't you simply say ", which is different from the present definition in Eq. (5), did not take into account the gravitational potential. Thus, the value of $\cdots$"?}

Here, we correct the effect of the gravitational potential energy in the data summarized in Ref.~\cite{kawaguchi2016jun} by employing the method introduced in Ref.~\cite{hayashi2021feb}, and recalibrate the ejecta-velocity fitting formula employing those corrected data as well as the results obtained in this paper. Our method to correct the ejecta velocity $v_{\rm eje,cor}$ is described in Appendix~\ref{sec:appendix2}. 
We note that, since the estimated value of the ejecta velocity for the models with small ejecta mass is not reliable, they employed only the models with $\hat{M}_{\rm eje}>0.003$. In this paper, we restrict the models in the same manner.
%\kk{(Did you also restrict the data obatined in this paper? May be it will be better to apply the same rule also for the present results.)}. 
%\kk{(It will be good to demonstrate in Appendix that this way of correction actually works. (It would be good to show an example in the same way as Kota presented in the meeting.))}

% By employing the corrected ejecta velocity $v_{\rm{eje,cor}}$ of the data summarized in Ref.\cite{kawaguchi2016jun},  we obtain the new calibration of the fitting formula:
% \begin{equation}\label{eq:fit_veje_cor}
%     v_{\rm{eje,cor}}=(0.01158Q+0.1469)c
% \end{equation}
%Both the original and the modified fitting formula don't fit well with $v_{\rm{eje}}$ in this paper.
% By employing the results of the ejecta velocity $v_{\rm{eje}}$ obtained in this paper, we obatin the fitting formula as:
% \begin{equation}\label{eq:fit_veje_new}
%     v_{\rm{eje}}=(0.02885Q+0.09501)c
% \end{equation}
%\kk{It is not clear how these formulas are drevied. We need to explain the method how you determined the parameters(breifly is fine.) and which data sets are used to derive each parameter of the formula.}
By employing all the data, including $v_{\rm{eje,cor}}$ in Ref.~\cite{kawaguchi2016jun} and $v_{\rm{eje}}$ in this paper, we obtain the fitting formula as
\begin{equation}\label{eq:fit_veje_all}
    v_{\rm{eje,fit}}=(0.01108Q+0.1495)c.
\end{equation}

\begin{figure}[t]
    \centering
    \includegraphics[width=0.45\textwidth]{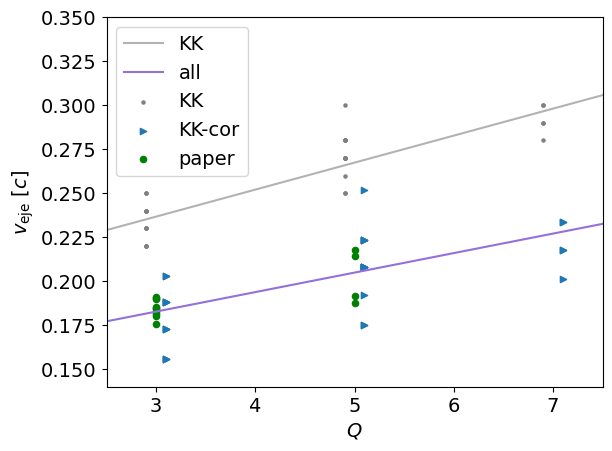}
    \caption{Original average ejecta velocity in Ref.~\cite{kawaguchi2016jun}, corrected velocity $v_{\rm{eje,cor}}$, and $v_{\rm{eje}}$ obtained in this paper as functions of mass ratio $Q$. ``KK'' denotes the data obtained from Ref.~\cite{kawaguchi2016jun}, ``KK-cor'' denotes the same data as ``KK'' but with the gravitational potential energy correction, $v_{\rm eje,cor}$, ``paper'' denotes the data newly obtained in this paper. The line with a label ``KK'' denotes the fitting formula obtained in the previous study~\cite{kawaguchi2016jun} (Eq.~(\ref{eq:fit_veje})). The line with the label ``all'' denotes the fitting formula obtained by employing both data of ``KK-cor'' and ``paper'' (Eq.~(\ref{eq:fit_veje_all})). The data points are slightly shifted in $x$-direction to make different sets of points separated.}
    \label{fig:veje_fit}
\end{figure}

Fig.~\ref{fig:veje_fit} shows that the average ejecta velocity in Ref.~\cite{kawaguchi2016jun} is generally larger than $v_{\rm eje}$ obtained in this paper, while after correction, $v_{\rm{eje,cor}}$ is generally closer to $v_{\rm eje}$. This indeed shows that the values of $v_{\rm eje,old}$ in Ref.~\cite{kawaguchi2016jun} are overestimated by neglecting the effect of the gravitational potential energy. 
%The slope of fitting formula of Eq.~\ref{eq:fit_veje_new} is larger compare to that of Eq.~\ref{eq:fit_veje} and Eq.~\ref{eq:fit_veje_cor}, which indicates that the mass ratio $Q$ has larger influence on $v_{\rm{eje}}$. However, since the number of models used in Ref.~\cite{kawaguchi2016jun} is far more than the models used in this paper (Eq.~\ref{eq:fit_veje_all}), the fitting formula obtained by employing all the data is almost the same as the fitting formula of Eq.~\ref{eq:fit_veje_cor}. 
The largest relative fitting error of the fitting formula is around $10\%$ for Eq.~(\ref{eq:fit_veje}) and is around $18\%$ for Eq.~(\ref{eq:fit_veje_all}). On the other hand, the relative numerical error of $v_{\rm{eje}}$ in this paper estimated from the different resolution runs is around $10\%$. This implies that there is a approximate linear relation between $v_{\rm{eje}}$ and $Q$. It may indicate that the ejecta velocity is not only determined by the mass ratio $Q$, but also influenced by other parameters, such as the BH spin, NS compactness, and initial spin misalignment angle. %For example, the models in Ref.~\cite{kawaguchi2016jun} has different initial misalignment angle between the BH spin and the orbital angular momentum, while for models in our paper, the angle is zero. Or it may be because the different configurations and numerical methods of simulations result in systemtically different results of the same models.

%The largest relative error of fitting formula is around $10\%$ for $v_{\rm eje,old}$ and $v_{\rm{eje}}$, and is around $20\%$ for $v_{\rm{eje,cor}}$ and for all data. Since the relative error of $v_{\rm{eje}}$ in this paper estimated by the difference of models of different resolutions is around $10\%$, the relative error of $20\%$ is too large, so there is no simple linear relations between $v_{\rm{eje}}$ and $Q$ if we consider all the data. It probaly means the ejecta velocity doesn't mainly depends on mass ratio $Q$, other parameters would also influence the results. For example, the models in Ref.~\cite{kawaguchi2016jun} has different initial misalignment angle between the BH spin and the orbital angular momentum, while for models in our paper, the angle is zero. Or it may be because the different configurations and numerical methods of simulations result in systemtically different results of the same models.

\begin{figure}[t]
    \centering
    \includegraphics[width=0.45\textwidth]{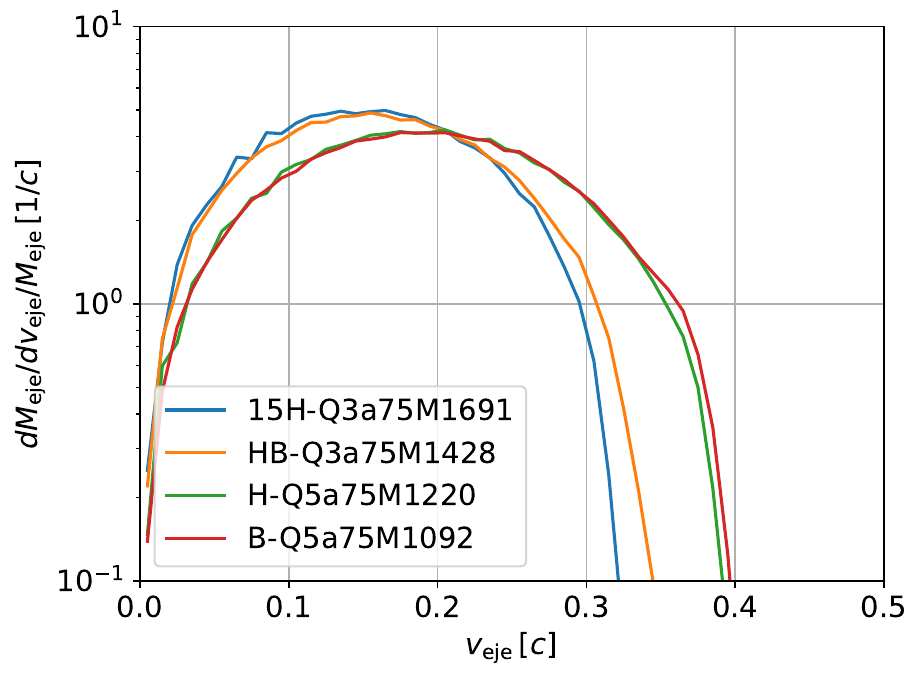}
    \caption{Ejecta velocity distributions normalized by the ejecta mass for selected models. 
    The models 15H-Q3a75M1691 and HB-Q3a75M1428 have $(Q, \chi_\mathrm{BH}, C_{\rm NS})=(3, 0.75, 0.182)$ in common, and the models H-Q5a75M1220 and B-Q5a75M1092 have $(Q, \chi_\mathrm{BH}, C_{\rm NS})=(5, 0.75, 0.147)$ in common.
    The models with the same values of $(Q, \chi_\mathrm{BH}, C_{\rm NS})$ approximately show the same distributions.}
    \label{fig:vel_dis}
\end{figure}

Fig.~\ref{fig:vel_dis} shows the velocity distributions of the ejecta normalized by its mass for selected models. The distributions are evaluated at $10$\,ms after the onset of the merger.
The models shown in the figure are 15H-Q3a75M1691 and HB-Q3a75M1428 which have $(Q, \chi_\mathrm{BH}, C_{\rm NS})=(3, 0.75, 0.182)$ in common, and H-Q5a75M1220 and B-Q5a75M1092 which have $(Q, \chi_\mathrm{BH}, C_{\rm NS})=(5, 0.75, 0.147)$ in common.
The distributions we obtained from our simulations show similar features to those obtained by Refs.~\cite{kyutoku2015aug,foucart2017jan,brege2018sep}; they peak at $v_{\rm eje} \approx 0.1c$--$0.3c$ and have a steep cut-off for both the lower and higher velocity sides.
%(but see the resolution study below).
The models with the same values of $(Q, \chi_\mathrm{BH}, C_{\rm NS})$ essentially show the same distributions having a peak at $v_{\rm eje} \approx 0.15c$ and $0.2c$ for the models with $(Q, \chi_\mathrm{BH}, C_{\rm NS})=(3, 0.75, 0.182)$ and $(5, 0.75, 0.147)$, respectively. The NS mass $M_{\rm NS}$ does not play a role in changing the result significantly.
%The only difference seen in the figure is that for the lower $M_{\rm NS}$ models (or the softer NS EOS models) the distribution tends to be extended to a slightly higher velocity side but it is not an outstanding feature. \kenta{Kiuchi: It may be better to explain why the lower $M_{\rm NS}$ models have a higher velocity. Presumably, the tidal disruption occurs closer to the BH, and consequently, the escape velocity becomes larger.} %Kota: This difference may be coming from the grid size difference so I deleted this sentence.

%%%%%%%%%%
%\addms{We also perform a careful resolution study in this paper fot the escape velocity distribution (see Fig.~\ref{fig:vel_dis_res}) and find a result worthy to be remarked. Fig.~\ref{fig:vel_dis_res} shows that for relatively low values of the escape velocity the convergence is achieved. However, for the high velocity side with $\agt 0.2c$ the convergence is quite slow. Specifically, for higher grid resolutions, more matter is present for the higher velocity side. Our interpretation for this is that with the higher grid  resolution the shock heating efficiency among the tidal tail (one-armed spiral arm of Fig.~\ref{fig:dens_profile}) is enhanced, and as a reult, more fraction of the matter gets the higher excape velocity. This suggests that in BH-NS mergers, the dynamical ejecta have two components; one is the component ejected by tidal force exerted at the tidal disruption process and the other is that ejected by the shock heating in the tidal tail, which is responsible for the higher velocity component. }

\begin{figure}[t]
    \centering
    \includegraphics[width=0.45\textwidth]{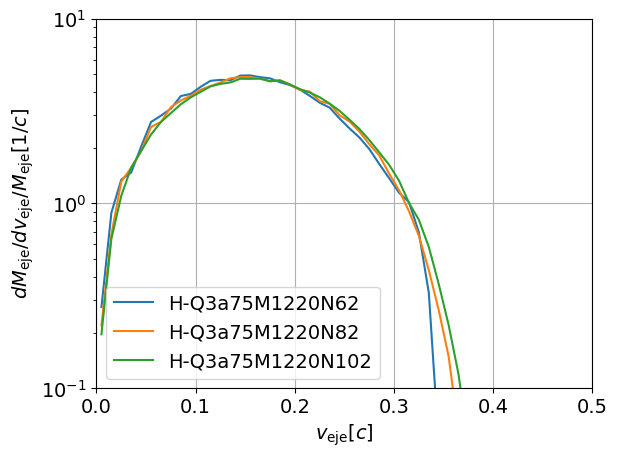}
    \caption{Ejecta velocity distributions normalized by the ejecta mass for model 15H-Q3a75M18 with resolutions N62, N82, and N102.
    For the higher resolutions, the distribution is extended to the higher velocity side.}
    \label{fig:vel_dis_res}
\end{figure}

We also perform a careful resolution study in this paper for the ejecta velocity distribution and find a result worthy of remark. 
Fig.~\ref{fig:vel_dis_res} shows the ejecta velocity distribution for model H-Q3a75M1220 with resolutions N62, N82, and N102.
It shows that for relatively low values of the ejecta velocity convergence is achieved. 
However, for the high-velocity side with $\agt 0.3c$ the convergence is not fully achieved. 
Specifically, for higher grid resolutions, more matter is present for the higher velocity side. 
Our interpretation for this is that with the higher grid resolution, the surface of the one-armed spiral, where the high-velocity ejecta exists, is better resolved and the pressure gradient plays a role in additionally accelerating the matter there.

%%%%%%%%%%

\subsubsection{Properties of gravitational waves}

\begin{figure}[t]
    \centering
    \includegraphics[width=0.45\textwidth]{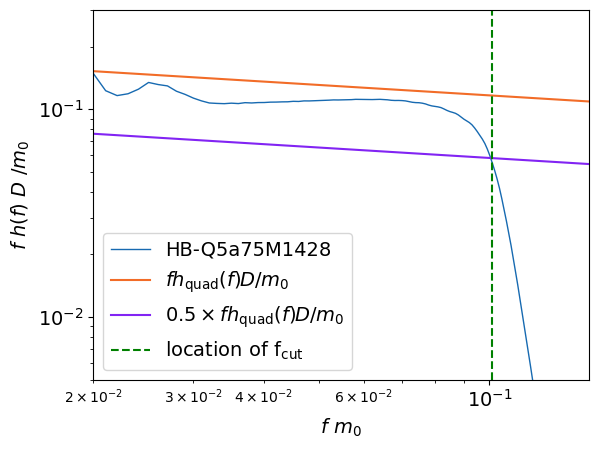}
    \caption{The example that shows how we determine the value of $f_{\rm cut}m_0$. The blue curve is the GW spectrum obtained from model HB-Q5a75M1428, and the purple line is half of the GW spectrum obtained from the quadrupole formula. We identify the intersection frequency as the value of the cutoff frequency $f_{\rm cut}m_0$.}
    %\kenta{Kiuchi: Please enlarge the font.}
    \label{fig:def_fcut}
\end{figure}

\begin{figure}[t]
    \centering
    \includegraphics[width=0.45\textwidth]{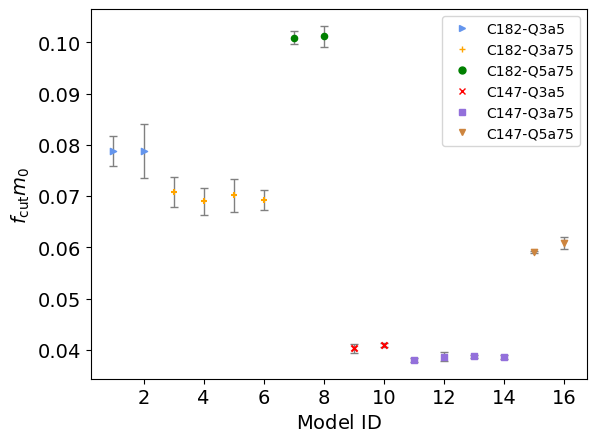}
    \caption{The normalized cutoff frequency $f_{\rm{cut}}m_0$ for models with the same values of $Q$, $\chi_{\rm{BH}}$, $C_{\rm{NS}}$, and different EOSs. The error bar of each model is taken to be the discrepancy of the results between $N=62$ and $N=82$ runs. Model ID corresponds to the column model ID in Tables~\ref{tab:C18} and \ref{tab:C147}.}
    %\kenta{"its error" in the first sentence is redundant. Please enlarge the font.}
    \label{fig:fixedC_fcut}
\end{figure}

%\wlh{Should the first few paragraphs and other introductions be moved to the diagnostics part? And shall we calculate the $f_{cut}$ of the mergers quantitively and put them in the table as we did for the remnant mass and ejecta mass? And I think we should look at the GW waveforms with fixed compactness (as the analysis in the last subsection). Simply concluding the behavior of cutoff frequencies with a, C, Q doesn't match the topic of this section, and it has been done by former studies. And Fig.~9, ~10 seem supposed to be replaced.}

%\kk{(!!!Explanations for the results are needed!!!)}

The location of the cutoff frequency in the amplitude of the GW spectrum at $f \agt 1$\,kHz contains rich information on the last stage of the inspiral phase and mainly depends on the orbital frequency of the tidal disruption% near the innermost stable circular orbit
\cite{Vallisneri2000,shibata2011aug,Pannarale2015}. If the NS is not tidally disrupted, the GW spectrum is characterized by the inspiral, merger, and ringdown waveforms, in which case the gravitational waveforms are nearly identical to those for binary BH mergers~\cite{Foucart2013Sep}. When tidal disruption happens, the wave amplitude quickly decreases before the ringdown phase, and the wave amplitude by the inspiral motion shuts down. This is because in the tidal disruptions, the NS matter becomes less dense and diffused, and the excitation of the quasinormal modes is suppressed by the phase cancellation~\cite{Nakamura1981,Shapiro1982,Nakamura1983,shibata2009feb}.
The GW spectra in tidal disruptions are characterized by their damping above the cutoff frequency, $f_{\rm{cut}}$.
%To quantify $f_{\rm{cut}}$, previous studies fitted the gravitational-wave spectra by a function of the form with seven free parameters $f_{\rm ins},f_{\rm dam},f_{\rm ins2},\sigma_{\rm %ins},\sigma_{\rm dam},\sigma_{\rm ins2},A$ and they defined $f_{\rm{cut}}$ as $f_{\rm{dam}}$ %\cite{shibata2009feb,kyutoku2010aug,kyutoku2011sep}:
%\begin{equation}
%    \begin{split}
%        \frac{fh_{\rm fit}(f)D}{m_0}=\frac{fh_{\rm 3PN}(f)D}{m_0}e^{-(f/f_{\rm ins})^{\sigma_{\rm ins}}}\\
%        +Ae^{-(f/f_{\rm dam})^{\sigma_{\rm dam}}[1-e^{-(f/f_{\rm ins2})^{\sigma_{\rm ins2}}}]}
%    \end{split}
%    \label{eq:fcutfit}
%\end{equation}
%where $h_{\rm 3PN}$ is the GW spectrum derived from the post-Newtonian theory.
To quantify $f_{\rm{cut}}$, previous studies fit the GW spectra by a function with seven parameters and define $f_{\rm{cut}}$ as one of the parameters \cite{shibata2009feb,kyutoku2010aug,kyutoku2011sep}. However, these two definitions are not robust because the fitting formula is not flexible enough.

For $f\rightarrow 0$, $f|h(f)|$ is proportional to $f^{-1/6}$ and can be approximated by the quadrupole formula $fh_{\rm quad}(f)$ with the point-particle approximation and the binary motion is determined by Newtonian gravity (e.g., Ref.~\cite{shibata2015textbook}). Therefore, we determine the value of $f_{\rm{cut}}$ to be the value of the intersection of the half of the GW spectrum obtained by the quadrupole formula and the one obtained from simulations; see Fig.~\ref{fig:def_fcut} as an example.

Fig.~\ref{fig:fixedC_fcut} shows the normalized cutoff frequency $f_{\rm cut}m_0$ and its error bars for models with the same values of $Q$, $\chi_{\rm{BH}}$, and $C_{\rm{NS}}$, and different EOSs. As in the cases of $M_{\rm rem}$ and $M_{\rm eje}$ (see Sec.~\ref{sec:fixed-C_mass}), the error bars of $f_{\rm cut}m_0$ are also taken to be the discrepancies of the results between $N=62$ and $N=82$ runs. 
By comparing the models with different values of $Q$ or $\chi_{\rm{BH}}$ or $C_{\rm{NS}}$, the figure shows that larger compactness $C_{\rm{NS}}$ and higher mass ratio $Q$ induce higher cutoff frequency, while higher BH spin $\chi_{\rm BH}$ induces lower cutoff frequency. These results are consistent with Eq.~(\ref{eq:R_dis_minus_R_isco}), since larger values of $C_{\rm{NS}}$ make the tidal disruption of the binary occur at a more inner orbit with a higher orbital frequency, higher values of $Q$ induce weaker tidal force in the vicinity of $R_{\rm ISCO}$, and higher values of $\chi_{\rm BH}$ gives larger spin-orbit repulsion, thus reduces the orbital angular velocity to maintain a circular orbit, decreases the radius of the ISCO, and therefore, makes NS more easily to be disrupted. These parameter dependence of the cutoff frequency shown in Fig.~\ref{fig:fixedC_fcut} are all consistent with the interpretation based on the comparison between $R_{\rm ISCO}$ and $R_{\rm dis}$ in Eq.~(\ref{eq:R_dis_minus_R_isco}).%, which implies that the naive the dependence holds for the case that the NS compactness is as high as $C_{\rm{NS}}\approx 0.182$.
%\kenta{Kiuchi: Question again. How good is the cuurent way of the error estimation?} 

It is also seen that within the numerical accuracy, the normalized cutoff frequency $f_{\rm cut}m_0$ for models with the same values of $Q$, $\chi_{\rm{BH}}$, and $C_{\rm{NS}}$ is approximately identical even though these models have different NS masses, i.e. different EOSs. The relative deviation between $f_{\rm cut}m_0$ within the same group is small, and as a quantity characterizing where tidal disruptions happen, the value of $f_{\rm{cut}}m_0$ depends solely on the ratio of $R_{\rm{dis}}$ and $R_{\rm{ISCO}}$. This dependence of $f_{\rm cut}m_0$ implies the relative location of tidal disruptions depends solely on $Q$,  $\chi_{\rm{BH}}$, and $C_{\rm{NS}}$.

\subsection{Massive NS case} \label{sec:mNS}

%\sout{As we \sout{discussed} \kk{demonstrated} in Sec.~\ref{sec:fixed-C}, $\hat{M}_{\rm{rem}}$ and $\hat{M}_{\rm{eje}}$ of the mergers \sout{mainly}\kk{primarily} depend on $C_{\rm{NS}}$, $a_{\rm{BH}}$, \kk{and} $Q$, and $M_{\rm{NS}}$ \kk{has only minor effect in the}\sout{does not affect the results a lot}. Therefore, it is reasonable to study the mergers of compact NSs by fixing their $M_{\rm{NS}}$ and changing their compactness through changing EOS.}
Table~\ref{tab:M18} shows the simulation results with a fixed NS mass $M_{\rm{NS}}=1.8M_{\odot}$ but with different EOSs. Massive NSs generally have weak or no tidal disruptions due to their high compactness, as indicated in our analysis in Sec.~\ref{sec:overview}.

\input{table_result_M18}
%\kk{(You need to explain more from which results (Table, Figure..) actually you could claim this statement. I think Figures and Tables should be tided up for this purpose.)}.
%We use the labels rem\_2012, rem\_2018, eje\_2016, eje\_2020 to denote the fitting formulas given in Refs.~\cite{foucart2012dec,foucart2018oct,kawaguchi2016jun,krger2020feb} respectively. 

To test the validity of fitting formulas derived in previous studies in more compact regions quantitatively, we calculate the $\chi$-square defined as 
\begin{equation}\label{eq:chi_square}
    \chi^2=\frac{1}{N_{\text{model}}-N_{\text{para}}}\sum_{i=1}^{N_{\text{model}}}\Delta_{i,\rm{fit}}^2,
\end{equation}
%\kenta{Kiuchi: Should $\Delta_i$ be $\Delta_{i,{\rm fit}}$? Or, $\Delta_i$ in Eq.~(26)?} 
where $N_{\text{model}}$ is the number of models in simulations, $\Delta_{i,\rm{fit}}$ is defined in Eq.~(\ref{eq:delta_fit}), and $N_{\text{para}}$ is the number of parameters used in the fitting formula.

$\Delta_{\rm{fit}}$ denotes the ratio of discrepancy of the results between the fitting formulas and the simulation data to the estimated numerical error of the simulation $\Delta_{\rm{NR}}$, which is defined by 
\begin{equation}
    \Delta_{\rm{fit}}= \frac{\hat{M}_{\rm{NR}}-\hat{M}_{\rm{fit}}}{\Delta_{\rm{NR}}}, \label{eq:delta_fit}
\end{equation}
where $\hat{M}_{\rm{NR}}$ and $\hat{M}_{\rm{fit}}$ denote the results from numerical simulations and the prediction of the fitting formulas, respectively, which can be either remnant or ejecta mass.
In Refs.~\cite{kawaguchi2016jun,krger2020feb}, for ejecta mass, $\Delta_{\rm{NR}}$ is assumed to be 
\begin{equation}\label{eq:error esti kyutoku}
    \Delta_{\rm{NR}}=\sqrt{(\hat{M}_{\rm{NR}}/10)^2+(1/50)^2},
\end{equation}
which holds for eje\_2016 and eje\_2020, while for remnant mass, Refs.~\cite{foucart2012dec,foucart2018oct} give 
\begin{equation}\label{eq:error esti rem}
    \Delta_{\rm{NR}}=\sqrt{(\hat{M}_{\rm{NR}}/10)^2+(1/100)^2},
\end{equation}
which is the case for rem\_2012 and rem\_2018.

%$\hat{M}$ in Eq.~\ref{eq:delta_fit} denotes the mass divided by the NS baryon mass, and 

\input{table_chi_square}

Since the fitting formulas were derived by minimizing the $\chi$-square in previous simulations, to achieve a benchmark of their accuracy, we first calculate the $\chi$-square of the fitting formulas with $\Delta_{\rm{NR}}$ from Eq.~(\ref{eq:error esti kyutoku}) for the ejecta mass and from Eq.~(\ref{eq:error esti rem}) for the remnant mass for numerical-relativity results given in Ref.~\cite{kyutoku2015aug}, of which parameters of BH-NS binaries are within the calibrated range. 
%The chi-square for the remnant mass are $\chi_{\rm{rem}}^{2012}=4.75$, and $\chi_{\rm{rem}}^{2018}=5.56$. The chi-square for the ejecta mass are $\chi_{\rm{eje}}^{2016}=0.0250$ and $\chi_{\rm{eje}}^{2020}=0.0601$. 
The results are shown in Table~\ref{tab:chi_square}. These results can be considered as typical values of the $\chi$-square. To evaluate how well the fitting formulas work for our simulations, we compare the $\chi$-square of our simulation data with that of Ref.~\cite{kyutoku2015aug}.%\cite{kawaguchi2015jul}. \kenta{Kiuchi: In where? Table VII compared the result in Ref. [24].}
%\kk{(!!!Please add some comments of the results of the benchmark test (ex., typical values, comparison with the typical numbers given in the reference (if available).)!!!)}
%$\chi_{\rm{rem}}^{2012},\chi_{\rm{rem}}^{2018},\chi_{\rm{eje}}^{2016},\chi_{\rm{eje}}^{2020}$ are 2.43, 4.16, 1.01, 0.00365. Moreover, for the C147 models, they are 1.24, 3.65, 0.119, 0.109 respectively; and for the C182 models, they are 1.23, 3.73, 0.108, 0.0825 respectively. 

Next, we calculate the $\chi$-square for our simulation data with $M_{\rm{NS}}=1.8M_{\odot}$, $C_{\rm{NS}}$ ranging from 0.194 to 0.252, and C182, C147 models %\kenta{Kiuchi: Again. C182 and C147 are already defined.}
with Eq.~(\ref{eq:chi_square}) and $\Delta_{\rm{NR}}$ given in Eqs.~(\ref{eq:error esti kyutoku}) and (\ref{eq:error esti rem}). 
%In general, the $\chi$-square for C182 and C147 models have similar values, which are different from the $\chi$-square for M18 models.
%For the remnant mass, the $\chi$-squares for M18, C147, and C182 for both fitting formulas are smaller than those of the models in Ref.~\cite{kyutoku2015aug} and close to unity, with $\chi^2$ for M18 larger than those of C182 and C147. 
For the remnant mass, the $\chi$-squares of M18, C147, and C182 for both fitting formulas are smaller than those of the models in Ref.~\cite{kyutoku2015aug}. For rem\_2012, its $\chi$-squares for our simulations are generally smaller than previous simulations and close to unity. $\chi_{\rm{rem}}^{2018}$ increases as the NS compactness decreases, and is generally larger than the corresponding $\chi_{\rm{rem}}^{2012}$.
%The decrease of $\chi^2$ of the remnant mass compared with models in Ref.~\cite{kyutoku2015aug} is probably because C182 and C147 models are in parameter space where the fitting formulas are well calibrated.
%\kk{(I think that the suppression of the numerical error in our simulation does not necessarily lead to the reduction of $\chi$-square. I guess this is simply due to the fact that C182 and C147 models are in the parameter space where the fitting formulas are well calibrated.)}.
%As to the ejecta mass, for C182 and C147 models, their $\chi$-square is slightly larger than those of models in Ref.~\cite{kyutoku2015aug}. 
As to the ejecta mass, for C182 and C147 models, their $\chi$-squares are smaller than those of models in Ref.~\cite{kyutoku2015aug}.
For the M18 case, its $\chi_{\rm{eje}}^{2016}$ is significantly larger than those of the other three cases, while its $\chi_{\rm{eje}}^{2020}$ is much smaller. 
Nevertheless, we find that within the numerical accuracy analyzed in previous studies, their fitting formulas for $\hat{M}_{\rm{rem}}$ and $\hat{M}_{\rm{eje}}$ work well even in the compact NS region since the $\chi$-squares are close to unity. %\kk{(!!!we need some reason why we can claim these formulas work well/not well (ex, because the $\chi$-square is close to the unity).!!!)} 
%While the $\chi$-squares for rem\_2012, rem\_2018, and eje\_2016 increase for more compact M18 models, $\chi^{2020}_{\rm{eje}}$ decreases as $C_{\rm{NS}}$ increases. 

%models in Ref.~\cite{kyutoku2015aug} and C182, C142, and its $\chi^{2020}_{\rm{eje}}$ is much smaller than that of models in Ref.~\cite{kyutoku2015aug} and C182, C142. 

%With the numerical error given in Eq.~\ref{eq:error esti kyutoku}, i.e. the error of previous simulations with lower resolutions, the results given by the fitting formulas for more compact NSs have almost the same $\chi$-square as that given for the $C_{\rm{NS}}<0.18$ region, meaning that their performance in the compact region is as good as that in the $C_{\rm{NS}}<0.18$ region.
It is not surprised to find that model rem\_2012 performs better than model rem\_2018 generally in the paper. The simulations used to fit rem\_2012 covers a relatively narrow parameter space compared to rem\_2018, and around 2/3 of them use $Q$=3--5 and $\chi_{\rm{BH}}^f$ = 0.50--0.75, consistent with the parameter range in our paper. Therefore, rem\_2012 is more likely to behave better when fitting the results from our simulations, while rem\_2018 is meant to avoid large errors on a larger scale with the trade-off of lower accuracy in some specific parameter range.

We then calculate the $\chi$-square of fitting formulas using $\Delta_{\rm{NR}}$ given by our simulations to test whether there could be room for improving the fitting formulas employing higher resolution data of a numerical-relativity simulation. In our research, the numerical error is estimated to be 
\begin{equation} \label{eq:error esti}
    \Delta_{\rm{NR}}=\sqrt{(\hat{M}_{\rm{NR}}/10)^2+(3/1000)^2}
\end{equation}
(see Appendix~\ref{sec:appendix1}). Eq.~(\ref{eq:error esti}) shows that the numerical errors of our simulations are smaller than those given by Eqs.~(\ref{eq:error esti kyutoku}) and~(\ref{eq:error esti rem}), and hence, the $\chi$-square calculated based on it gives more severe examination to the fitting formula.

Results are shown in Table~\ref{tab:chi_square}. The $\chi$-squares for M18 are significantly larger than those of C147 and C182 except for eje\_2020, which are much smaller. 
%The discrepancies of $\chi^2$ between C182 and C147 become more prominent and $\chi^2$ of C182 is larger. 
The reason for this increase is that mergers with compact NSs result in smaller values of $\hat{M}_{\rm{rem}}$ and $\hat{M}_{\rm{eje}}$ and therefore sensitive to the lower limit of the estimated numerical error. Assuming a smaller numerical error enhances their $\chi$-squares significantly. $\chi^{2018}_{\rm{rem}}$ of C182, C147 also increases significantly, while other $\chi-$squares for C147 and C182 models remain close to the unity.

We find that some $\chi$-squares in Table~\ref{tab:chi_square} are much less than unity. This is because the numerical error we use is the rough upper estimation, which may be overestimated especially when $\hat{M}_{\rm{NR}}$ is small, and makes $\chi$-squares small.
%We find that for $M_{\rm{NS}}=1.8M_{\odot}$ cases, $\chi_{\rm{rem}}^{2012},\chi_{\rm{rem}}^{2018},\chi_{\rm{eje}}^{2016},\chi_{\rm{eje}}^{2020}$ are 17.5, 42.8, 44.8, 0.162 respectively. While for the C147 models, they are 1.03, 3.66, 2.87, 1.40 respectively; and for the C182 models, they are 2.23, 23.0, 4.53, 3.50 respectively, which are much smaller than that of M18 models except eje\_2020.

Therefore, if we assume a smaller numerical error of our simulations, rem\_2012, rem\_2018, and eje\_2016 have room for improvement, particularly for compact NS cases. Their $\chi$-squares increase as the NS compactness increases, and for the M18 case, their $\chi$-squares are significantly larger. This is reasonable because these fitting formulas are calibrated in the parameter space of $C_{\rm{NS}}<0.18$. Only $\chi_{\rm{eje}}^{2020}$ decreases as the compactness increases and has a better performance in the compact NS region. Thus, if we want to estimate the remnant disk mass for $C_{\rm{NS}}>0.194$ at higher accuracy, existing fitting formulas require improvement. On the other hand, the fitting formula of the ejecta mass given in Ref.~\cite{krger2020feb} still works well in this region even if we assume a smaller numerical error of our simulations since its $\chi$-square is less than the unity. In addition, if we require a higher accuracy of the estimation of the remnant mass, rem\_2012 works better than rem\_2018.%\kk{(!!!we need some reason why we can claim Ref.~\cite{krger2020feb} works well (ex, because the $\chi$-square is close to the unity.!!!)}.
%\kenta{Kiuchi: Maybe, you can comment that rem\_2018 does not improve rem\_2012 generally.}
% \begin{figure}
%     \centering
%     \includegraphics[width=\linewidth]{fig_rem_M18.png}
%     \caption{Simulation results for $M_{\rm{NS}}=0.18M_{\odot}$. The simulation results are denoted by the cross markers, the predictions of rem\_2012 are denoted by point markers, and the predictions of rem\_2018 are denoted by triangle markers. The error bar is the discrepancies between N62 and N82 simulations.}
%     \label{fig:rem_M18}
% \end{figure}

%Fig.~\ref{fig:rem_M18} shows the error bar of the remnant mass and the prediction given by the fitting formula from Re.~\cite{foucart2012dec,foucart2018oct}.

% \begin{figure*}
%     \centering
%     \includegraphics[width=2\columnwidth]{rem_M18.png}
%     \caption{Rem\_M18, solid li
%ne:q5a75, dot-dashed line: q3a75, dashed line: q3a5}
%     \label{fig:rem_M18}
% \end{figure*}

%\kk{(!!!Results for small compactness models are needed!!!)}

\section{Conclusion \& Discussion} \label{sec:conclusion}

\begin{figure*}[t]
    \centering
    \includegraphics[width=\linewidth]{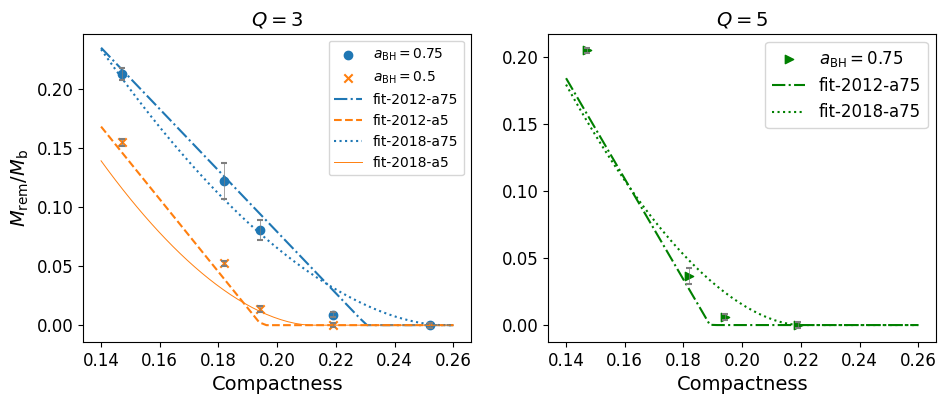}
    \caption{Normalized remnant mass as a function of the compactness. For models with the same values of $Q$, $\chi_{\rm{BH}}$, and $C_{\rm{NS}}$, since the discrepancies between them are within the margin of error, the simulation results are taken randomly from one of them, and the errorbar is taken to be the largest error range of them.}
    \label{fig:CNS-Mrem}
\end{figure*}

% to test the validity of fitting formulas with more compact NSs, as well as the consistency of $\hat{M}_{\rm{rem}}$ and $\hat{M}_{\rm{eje}}$ \wlh{(and $v_{eje}.f_{cut}$)} given the same $C_{\rm{NS}}, \chi_{\rm{BH}}, Q$ at a higher precise.
%We perform numerical relativity simulations for BH-NS mergers with fixed NS compactness $C_{\rm{NS}}=0.182, 0.147$, and compact NSs with three EOS, where $M_{\rm{NS}}=1.8M_{\odot}$.
We perform numerical-relativity simulations for various setups of BH-NS mergers with $ Q=3, 5$, $\chi_{\rm{BH}}=0.5, 0.75$, and $C_{\rm{NS}}=0.147\mbox{--}0.252$ varying the NS mass. Our results with the fixed NS compactness $C_{\rm{NS}}=0.182$ and $0.147$ show that given the identical values of $Q$, $\chi_{\rm{BH}}$, and $C_{\rm{NS}}$, $\hat{M}_{\rm{rem}}$ and $f_{\rm{cut}}m_0$ agree with each other within the numerical accuracy of our study. We also found that this is the case for $\hat{M}_{\rm{eje}}$, except for only one data set which is slightly out of the numerical accuracy. Therefore, the hypothesis that $\hat{M}_{\rm{rem}}$, $\hat{M}_{\rm{eje}}$, $v_{\rm{eje}}$ and $f_{\rm{cut}}m_0$ depend only on $Q$, $\chi_\mathrm{BH}$, and $C_{\rm{NS}}$ is approximately true for the NS mass in a wide range of $1.1\mbox{--}1.7\,M_\odot$. This justifies the approach of studying the dependence of NS tidal disruptions on the NS compactness by fixing the NS mass but changing the EOS, which is employed in the previous studies.

%Therefore, these three physical quantities are sufficient for us to fit the $\hat{M}_{\rm{rem}}$, $\hat{M}_{\rm{eje}}$ and $f_{\rm{cut}}$.

%This conclusion has been reached through previous studies, and we verify it at higher precision, making it valid to study the tidal disruptions of NSs with different compactness by fixing their mass and changing their EOS.
%\wlh{(Conclusions for fcut. I guess $f_{cut}$ behaves in the same way as $M_{rem}$ because they are all determined by TD.)}

We then performed numerical-relativity simulations of BH-NS mergers for a large value of the NS mass ($M_{\rm NS}=1.8\,M_\odot$), and examined the accuracy of the previous fitting formulas of $\hat{M}_{\rm{rem}}$, $\hat{M}_{\rm{eje}}$, and ejecta velocity $v_{\rm{eje}}$ for the system with a large value of the NS compactness ($C_{\rm NS}\ge 0.19$). As to the four fitting formulas of $\hat{M}_{\rm{rem}}$ and $\hat{M}_{\rm{eje}}$, we find that %except eje\_2016\kk{(for me, eje\_2016 seems to be ok because the $\chi$-square is still within unity, which means the error is within the expectation.)}, others 
they still work well for compact NSs within the error range of their studies. However, if we require a higher-precision prediction, the fitting formulas of the remnant mass could give inaccurate predictions for $C_{\rm{NS}}>0.194$ cases, leaving room for further improvements, and rem\_2012 is generally more accurate than rem\_2018.
%\sout{, and thus required to be improved}\kk{(!!!The fact that there is room for improving the fitting formulas does not necessarily mean we should: It also depends on whether it is needed or not.!!!)}. 
As to the ejecta mass, the fitting formula eje\_2020 from Ref.~\cite{krger2020feb} still works well for more massive NSs with $C_{\rm{NS}}$ between 0.194 and 0.252.

For the case of ejecta velocity, we corrected the data of average ejecta velocity in Ref.~\cite{kawaguchi2016jun} to take into account the gravitational potential effect, and recalibrated the ejecta-velocity fitting formula employing these corrected data as well as the results obtained in our present simulations. We obtained the simple linear function as the fitting formula between $v_{\rm eje}$ and $Q$ as in Ref.~\cite{kawaguchi2016jun}. We found that the predicted ejecta velocity is systematically smaller than those of the previous studies. The relative fitting error of the fitting formula is around $10\%$ for the data in this paper and in Ref.~\cite{kawaguchi2016jun}, and grows to $18\%$ if we use all the data to obtain a new fitting formula. It is larger than $10\%$ of the relative numerical error of $v_{\rm eje}$ estimated from different resolutions runs. This indicates that the ejecta velocity has a dependence on the other parameters which cannot be captured by the simple linear relations between $v_{\rm eje}$ and $Q$.
%\kk{(Conclusion from the ejecta velocity analysis should be also added in this section.)}
%We found $v_{\rm eje}$ increases as $Q$ get larger, and depends weakly on other parameters.

Our results give some guidance on how to improve the fitting formulas for the remnant mass. From Fig.~\ref{fig:CNS-Mrem}, which shows $\hat{M}_{\rm{rem}}$ as a function of $C_{\rm{NS}}$, we can see that rem\_2012 works best for Q3a5 models, even for compact NSs. The inaccuracy of the fitting formula is caused mainly by the large discrepancies between the fitting and the simulation results for Q3a75 models. The fitting formula generally overestimates the remnant mass for all Q3a75 models, but this overestimation is only significant for mergers with small remnant masses, i.e. with compact NSs. For example, H-Q3a75 has only 0.9\% of its baryon mass remaining outside the BH after the merger, and the rem\_2012 gives a prediction of 2.8\%. Rem\_2018 overestimates the remnant mass for compact NSs and is generally less accurate than rem\_2012. Therefore, a linear function of $C_{\rm{NS}}$ is still a good choice for the fitting formula of the remnant mass and the second term in Eq.~(\ref{eq:rem_2012}) requires modification to get more accurate results. 
%This overestimation can lead to a wrongful judgment of criteria at the edge of tidal disruptions. \kenta{Kiuchi: It is not easy for me to understand this sentence. Does "a wrongful judgment" mean that if we judge whether or not the tidal disruption occurs in observational data?} 

The improvement in the fitting formulas in such a parameter region may be not very important at this moment, since the overestimation is only significant for the cases having weak tidal disruptions, for which only faint EM counterparts can associate with the GW signals and are less likely to be detected by observations. However, the improvement in such a parameter region can play an important role in interpreting the observational data once faint EM counterparts are detected or deep EM upper-limits are obtained in the future BH-NS GW events (cf, GW190425 and GW190814~\cite{Kyutoku:2020xka,Kawaguchi:2020osi}).

%However, since the overestimation is only significant for models having weak tidal disruptions, and running simulations all over again is resource-consuming, the modification is not very important at the moment. This is because mergers with weak tidal disruptions have small $M_{\rm{rem}},M_{\rm{eje}}$, and can generate faint EM counterparts, which are less likely to be detected by observations.

%\kk{(It is good to have some comments what is the impact of improving the fitting formula in such parameter regions: Is it important to improve the parameter region where anyway the disk and ejecta masses are small (=faint EM counterparts)? Because we may need to rerun all the simulations for the entire parameter ranges to improve the fitting formula, and it's quite computationally expensive.)}

%Therefore, there is still space for improvement in fitting formulas of the remnant mass for more compact NSs. 

\begin{acknowledgments}
We would like to thank Francois Foucart for his helpful comments. Numerical computation was performed on Yukawa21 at Yukawa Institute for Theoretical Physics, Kyoto University and the Yamazaki, Sakura, Cobra, and Raven clusters at Max Planck Computing and Data Facility. This work was supported by Grant-in-Aid for Scientific Research (20H00158, 23H04900, and 23H01172) of JSPS/MEXT. Shichuan Chen is supported by China Scholarship Council.
\end{acknowledgments}

\appendix

\section{ESTIMATION OF ERROR CAUSED BY GRID RESOLUTION}\label{sec:appendix1}

%As described in Sec \ref{sec:fixed-C}, we perform simulations with fixed $a_{\rm{BH}}, Q, C_{\rm{NS}}$ and change the EOS of NS, to see whether the hydrodynamic quantities $M_{\rm{rem}}/M_{\rm{b}}, M_{\rm{eje}}/M_{\rm{b}}$ only depend on these three parameters. To determine whether the discrepancies between results from different models are from physical processes or numerical errors, we have to estimate errors and uncertainties in our simulations. 

\begin{figure*}[t]
    \centering
%    \subcaptionbox{Errors of $M_{\rm{rem}}/M_{\rm{b}}$}{\includegraphics[width = 0.4\textwidth]{fig_rem_N62_N82.png}}
%    \subcaptionbox{Errors of $M_{\rm{eje}}/M_{\rm{b}}$}{\includegraphics[width = 0.4\textwidth]{fig_eje_N62_N82.png}}
%    \hfill
\includegraphics[width = 0.4\textwidth]{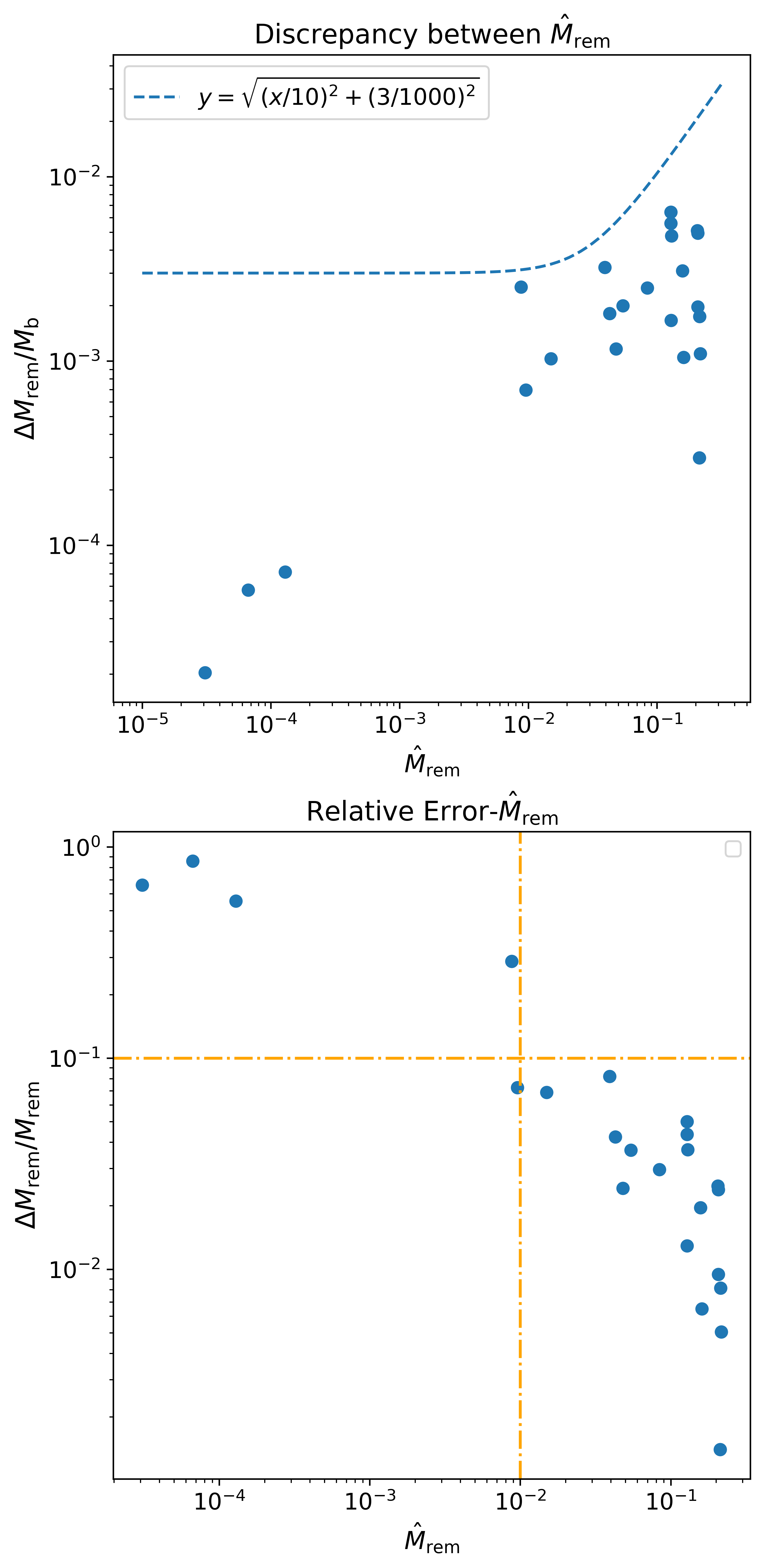}~~~
\includegraphics[width = 0.4\textwidth]{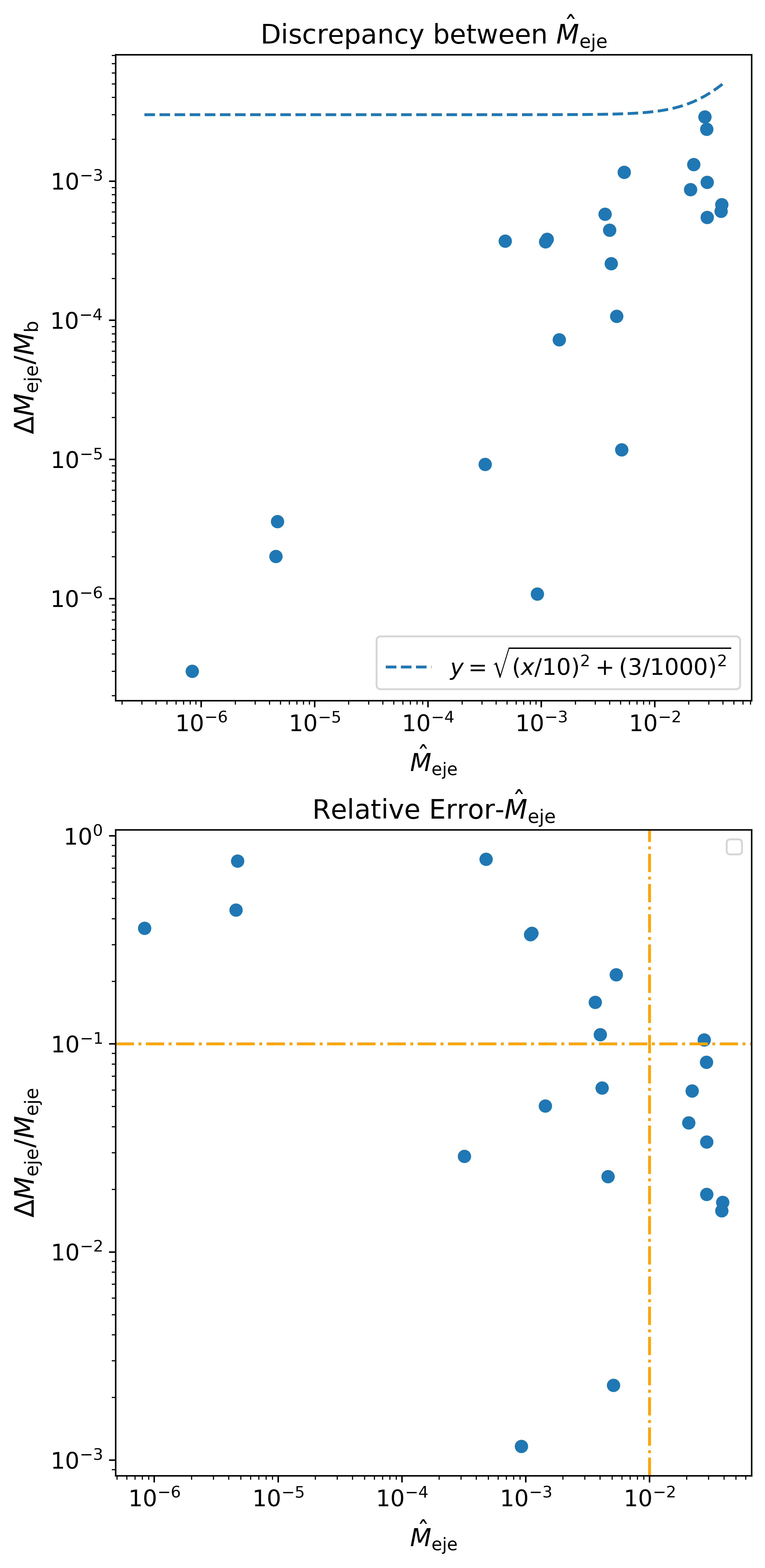}
    \caption{Discrepancy of $M_{\rm{rem}}/M_{\rm{b}}$ and $M_{\rm{eje}}/M_{\rm{b}}$ 
        between runs with different grid resolutions. The $x$-axis is taken to the results from $N=82$ runs. The scatters are taken to be the minimum discrepancy among N62, N82, and N102 runs. The blue-dashed curves are an upper estimation of the numerical error of the normalized remnant and ejecta mass, and the orange-dashed lines are an estimation of the relative numerical error.
    %\kenta{Kiuchi: $M_{\rm{rem}}/M_{\rm{b}}$ and $M_{\rm{eje}}/M_{\rm{b}}$ should be $\hat{M}_{\rm rem}$ and $\hat{M}_{\rm eje}$. The explanation for the green dashed curves and orange dashed lines is necessary. The titles of the bottom panels are wrong. It should be "Relative error-$M_{\rm{rem}}$". Please enlarge the font.}
    }
    \label{fig:error}
\end{figure*}
    
The mass of the remnant disk and the ejecta are only a fraction of the NS mass, and thus, can have large errors in the numerical results. Here we estimate the errors due to finite grid resolutions, as shown in Fig.~\ref{fig:error}. Both masses depend sensitively on the grid resolution. %For $\hat{M}_{\rm{rem}}$, the relative error ranges from 0.6\% to 50\%. For the $\hat{M}_{\rm{eje}}$, the relative error ranges from 0.1\% to 50\%.

 As shown in the lower plots, for $\hat{M}>0.01M_\odot$, the relative errors are generally less than 10\%, while for very small values of $\hat{M}$, the relative error can be up to 100\%. We expect that it is possible to obtain a function for the numerical errors, analogous to that given in Eq.~(\ref{eq:error esti kyutoku}) and Eq.~(\ref{eq:error esti rem}). Therefore, the coefficient in front of $\hat{M}$ in Eq.~(\ref{eq:error esti}) can be taken as 0.1, and the constant part is taken as 0.003. As shown in Fig.~\ref{fig:error}, the error for $\hat{M}_{\rm{eje}}$ and $\hat{M}_{\rm{rem}}$ can be described using a uniform estimation as it is in Eq.~(\ref{eq:error esti}).%\kenta{Kiuchi: It is hard to understand this sentence. Do you mean the coefficient in front of $\hat{M}$ in Eq. (29)?}

\input{table_convergence}
We list numerical results for some models with different grid resolutions $N=62,\ 82$ and $102$ in Table~\ref{tab:convergence}. 
%, and plot the time evolution of the remnant disk mass for different grid resolutions in Fig.~\ref{fig:convergence_of_rem}, the time evolution of the ejecta mass for different grid resolutions in Fig.~\ref{fig:convergence_of_eje}. 
The relative error between N82 and N102 runs ranges from 0.9\% to 3\% for $\hat{M}_{\rm{rem}}$, and ranges from 1\% to 34\% for $\hat{M}_{\rm{eje}}$. 
%\kenta{Kiuchi: The latter sentence sounds the error estimation is not stringent or reliable because the dashed curve in the upper-right panel in Fig. 12 tells the estimated error is O(0.1)\%.}
%\wlh{Luohan: The dashed curve is $\Delta M_{eje}/M_b$, and the relative error is $\Delta M_{eje}/M_{eje}$, which is between 0.1\% and 10\% for N82 and N102, as shown in the lower-right panel in Fig. 12.}
The relative error of the ejecta velocity between $N=82$ and $N=62$ runs ranges from 2\% to 5\%. The discrepancies of $f_{\rm{cut}}m_0$ between $N=62$ and $N=82$ range from 0.4\% to 5\%. The discrepancies of the spin are around 0.1\% for the three resolutions. In general, due to the complexity of hydrodynamics, hydrodynamic quantities have larger uncertainties.
%The discrepancies of numerical results between different resolutions decrease as $N$ increases, although systemic convergence properties are not seen. 
$\hat{M}_{\rm{rem}}$ decreases as the grid resolution increases, while systematic convergence properties of other quantities are not seen, and the discrepancies of numerical results among different resolutions do not decrease necessarily as the grid resolution increases.
The reason for this unsystematic behavior is likely that (i) in the presence of shocks (e.g., in the tidal tail), numerical accuracy becomes the first-order convergence and hence the error can be enhanced in an unpredicted manner. (ii) the motion of the matters is affected by the background atmosphere in the simulations~\cite{kyutoku2015aug}, the error associated with the artificial atmosphere decreases significantly as grid resolutions are improved, due to the suppression of spurious shocks at the stellar surface.

% \begin{figure}[!htbp]
%     \centering
%     \includegraphics[width=0.45\textwidth]{fig_convergence_of_rem.png}
%     \caption{Comparations of disk mass divided by the NS baryon mass evolution for different grid resolutions N62, N82, N102. Solid lines stand for N62, dot-dashed lines for N82, and dotted lines for N102.}
%     \label{fig:convergence_of_rem}
% \end{figure}

% \begin{figure}[!htbp]
%     \centering
%     \includegraphics[width=0.45\textwidth]{fig_convergence_of_eje.png}
%     \caption{Comparations of ejecta mass divided by the NS baryon mass evolution for different grid resolutions N62, N82, N102. Solid lines stand for N62, dot-dashed lines for N82, and dotted lines for N102.}
%     \label{fig:convergence_of_eje}
% \end{figure}

\section{CORRECTION OF EJECTA VELOCITY}\label{sec:appendix2}

\begin{figure}[t]
    \centering
    \includegraphics[width=0.45\textwidth]{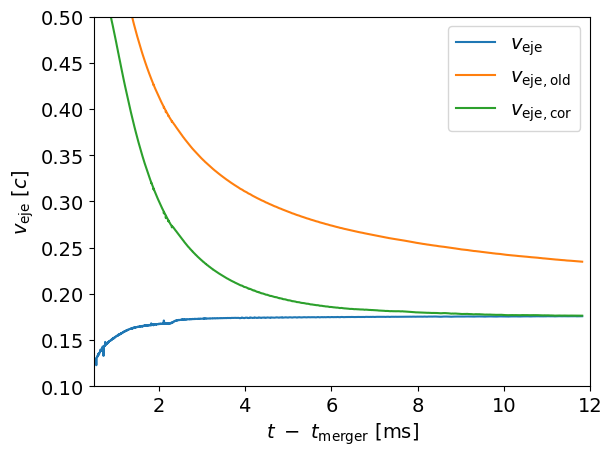}
    \caption{Time evolution of $v_{\rm eje}$ for model 15H-Q3a75M1691 with different definitions. Here, the results are from $N=82$ runs. $v_{\rm eje}$ is defined by Eq.~(\ref{eq:fit_veje}), $v_{\rm eje,old}$ by Eq.~(\ref{eq:veje_old}), and $v_{\rm eje,cor}$ by Eq.~(\ref{eq:veje_cor}).}
    %\kenta{Kiuchi: Please enlarge the font.}
    \label{fig:veje_cor}
\end{figure}

Different from Eq.~(\ref{eq:T_eje}), the kinetic energy of the ejecta defined in Ref.~\cite{kawaguchi2016jun} is
\begin{equation}
    T_{\rm eje,old}:= E_{\rm eje} - U_{\rm eje} - M_{\rm eje}. 
    \label{eq:T_eje_old}
\end{equation}
The total energy of the ejecta $E_{\rm eje}$ is defined by
\begin{equation}
     E_{\rm eje}:=\int_{-u_t>1,r>r_{\rm AH}}\rho_* \hat{e}\,d^3x.
    \label{eq:E_eje}
\end{equation}
The internal energy of the ejecta $U_{\rm eje}$ is defined by
\begin{equation}
     U_{\rm eje}:=\int_{-u_t>1,r>r_{\rm AH}}\rho_* \epsilon\, d^3x.
    \label{eq:U_eje}
\end{equation}
The mass of ejecta $M_{\rm eje}$ is defined as in Eq.~(\ref{eq:Meje}). The definitions of $\rho_*$, $\hat{e}$ and $\epsilon$ are listed in Table.~\ref{tab:quantity}.

Assuming the Newtonian dynamics, the averaged velocity of the ejecta can be evaluated as
\begin{equation}
    v_{\rm eje,old}:=\sqrt{\frac{2T_{\rm eje,old}}{M_{\rm eje}}}. 
    \label{eq:veje_old}
\end{equation}

The gravitational potential energy is not excluded in $T_{\rm eje,old}$, and $v_{\rm eje,old}$ is overestimated. To subtract the effect of this gravitational potential energy, we can approximately correct the velocity as in Ref.~\cite{hayashi2021feb}

\begin{equation}
    v_{\rm eje,cor}:=\sqrt{v^2_{\rm eje,old}-\frac{2m_0}{v_{\rm eje,old}(t-t_{\rm merger})}},
    \label{eq:veje_cor}
\end{equation}
where $v_{\rm eje,old}$ is evaluated at time $t$.

Fig.~\ref{fig:veje_cor} shows the time evolution of $v_{\rm eje}$ with different definitions for a representative model 15H-Q3a75M1691. If we choose $v_{\rm eje}$ defined in Eq.~(\ref{eq:v_eje}) to be the fiducial value, we find that $v_{\rm eje,old}$ is still overestimated by $\sim 50\%$ at 12\,ms after the onset of the merger. The corrected velocity of ejecta $v_{\rm eje,cor}$ is closer to $v_{\rm eje}$ compared to $v_{\rm eje,old}$, and asymptotically approaches $v_{\rm eje}$ at 10\,ms after the onset of the merger. Thus, it is reasonable to use Eq.~(\ref{eq:veje_cor}) to correct previous data of the average velocity of ejecta.

\bibliography{paper}
\end{document}

%% file: table_physical_quantities.tex
\begin{table}[!htbp]
    \centering
%    \resizebox{\linewidth}{!}{    
\caption{Our convention of notation for physically important quantities, geometric variables, and hydrodynamic variables.}
    \label{tab:quantity}
    \begin{tabular}{cc}
    \hline
    Symbol & Description\\
    \hline
    $M_{\rm{BH}}$&Gravitational mass of the black hole in isolation\\
    $\chi_{\rm{BH}}$&Dimensionless spin parameter of the black hole\\
    $M_{\rm{NS}}$& Gravitational mass of the neutron star in isolation\\
    $R_{\rm{NS}}$& Circumferential radius of the neutron star in isolation\\
    $M_{\rm{b}}$& Rest baryon mass of the neutron star in isolation\\
    $C_{\rm{NS}}$& Compactness of the neutron star: $C_{\rm{NS}}=M_{\rm{NS}}/R_{\rm{NS}}$\\
    $Q$& Mass ratio of BH-NS: $Q=M_{\rm{BH}}/M_{\rm{NS}}$\\
    %$R_{\rm{ISCO}}$& Innermost stable circular orbit of the black hole\\
    $M_{\rm{rem}}$& Remnant disk mass after the merger\\
    $M_{\rm{eje}}$& Ejecta mass after the merger\\
    $\hat{M}$& Mass normalized by $M_{\rm{b}}$\\
    $m_0$& Initial total gravitational mass of the system\\
    &at infinite separation\\
    \hline
    $\gamma_{ij}$& Induced metric on the $t$ = const hypersurface\\
    $\alpha$&Lapse function\\
    $\beta^i$&Shift vector\\
    $\gamma$&Determinant of $\gamma_{ij}$\\
    \hline
    $\rho$&Baryon rest-mass density\\
    $u^{\mu}$&Four velocity of the fluid\\
    $P$&Pressure\\
    $\epsilon$&Specific internal energy\\
    $\rho_*$& Conserved baryon rest mass density: $\rho \alpha\sqrt{\gamma}u^t$\\
    $\Gamma$&Adiabatic index\\
    $h$&Specific enthalpy: $1 + \epsilon + P/\rho$\\
    $\hat{e}$&Specific energy: $h\alpha u^{t} - P/\rho \alpha u^{t}$\\
    \hline
    \end{tabular}
%    }
\end{table}

%% file: table_EOS.tex
\begin{table}[t]
    %\resizebox{\linewidth}{!}{
\caption{Piecewise polytropic EOS Models employed in the present simulations. Model EOS name, $P_{\rm fid}$: the pressure at the fiducial density $\rho_{\rm fid}=10^{14.7}\rm{g\,cm^{-3}}$, $\Gamma_i$ ($i=1$--3), and the maximum gravitational mass of spherical NSs for the given EOS. The first four EOSs are composed only of two peices.} %\textcolor{cyan}{Kiuchi: $\rho_2$ and $p_1$ should be $\rho_{\rm fid}$ and $P_{\rm fid}$, respectively.}}
\begin{tabular}{c|ccccc}
    \hline
    EOS & $\log_{10}P_{\rm fid}[\rm{dyne/cm^2}]$& $\Gamma_1$ & $\Gamma_2$ & $\Gamma_3$ & $M_{\rm{max}}[M_{\odot}]$\\
    \hline
    15H& 34.700 &3.000 & N/A& N/A& 2.53\\ %2.525 \\
    H  & 34.500 &3.000 & N/A& N/A& 2.25\\ %2.249 \\
    HB & 34.400 &3.000 & N/A& N/A& 2.12\\ %2.122 \\
    B  & 34.300 &3.000 & N/A& N/A& 2.00\\ %2.003 \\
    \hline
    H4 & 34.669 & 2.909 &2.246 &2.144 &2.03\\
    APR4 & 34.269 &2.830 &3.445 &3.348 &2.20\\
    \hline
\end{tabular}
    %}
    \label{tab:EOS}
    \end{table}

%% file: table_initial_para.tex
\begin{table*}[t]
\caption{Parameters of initial data and grid structure for all the models studied in this paper. The model name contains the EOS employed, the mass ratio ($Q$), and the dimensionless spin of the black hole $\chi_{\rm BH}$. $M_{\rm {NS}}, M_{\rm {NS}}^{\rm{b}}, R_{\rm{NS}}, C_{\rm {NS}}$, and $M_{\rm BH,0}$ are the initial NS mass, NS restmass, NS radius, NS compactness, and BH mass, respectively.
$m_0\Omega_0$ is the dimensionless initial orbital angular velocity of the system. %$l_c,l_f$ are the numbers of nonmoving grids and a half of finer moving grids. $J_0$ is the orbital angular momentum.
$L$ is the box size of the simulation, $\Delta x$ is the grid spacing at the finest level, and $R_{\rm diam}/\Delta x$ is the grid number within the semimajor diameter of the NS. $N$ denotes the grid resolution of the simulation. The table contains three parts. The first part shows the parameters for the fixed value of the NS mass as  $M_{\rm{NS}}=1.8M_{\odot}$, the second and third parts show the parameters for the fixed values of $C_{\rm{NS}}=0.182$ and  $0.147$, respectively. $\Delta x$ and $R_{\rm{diam}}/\Delta x$ are calculated with the grid resolution of $N=82$.}
\label{tab:initial_para}
\resizebox{\linewidth}{!}{
\begin{tabular}{c|ccccccccccccc}
\hline
Model &EOS &$M_{\rm NS}[M_{\odot}]$ &$M_{\rm {NS}}^{\rm{b}}[M_{\odot}]$ &$R_{\rm{NS}}$[km]&$C_{\rm NS}$ &$Q$ &$M_{\rm BH,0}[M_{\odot}]$ &$\chi_{\rm BH}$ &$m_0\Omega_0$ %&$J_0/M_{\odot}^2$ %&$l_c$ &$l_f$ 
&$\Delta x[\rm m]$ &$R_{\rm diam}/\Delta x$ &$L[\rm km]$ &$N$\\

\hline
15H-Q3a5M18 &15H &1.800 &2.022 &13.70 &0.194 &3.0 &5.400 &0.50 &0.0377 &204 & 134& 8553&62, 82\\

H-Q3a5M18 &H &1.800 &2.056 &12.14 &0.219 &3.0 &5.400 &0.50 &0.0377 &180 & 135& 7565&62, 82\\

%\specialrule{0em}{1pt}{1pt}
15H-Q3a75M18 &15H &1.800 &2.022 &13.70 &0.194 &3.0 &5.400 &0.75 &0.0375 &204 & 134& 8553&62, 82, 102\\

H-Q3a75M18 &H &1.800 &2.056 &12.14 &0.219 &3.0 &5.400 &0.75 &0.0375 &180 & 135& 7565&62, 82\\

B-Q3a75M18 &B &1.800 &2.103 &10.55 &0.252 &3.0 &5.400 &0.75 &0.0375 &163 & 129& 6842&62, 82\\

%\specialrule{0em}{1pt}{1pt}
15H-Q5a75M18 &15H &1.800 &2.022 &13.70 &0.194 &5.0 &9.000 &0.75 &0.0445 &203 & 134& 8523&62, 82\\

H-Q5a75M18 &H &1.800 &2.056\ &12.14 &0.219 &5.0 &9.000 &0.75 &0.0445 &180 & 135& 7565&62, 82\\

\hline
15H-Q3a5M1691 &15H &1.691 &1.884 &13.72 &0.182 &3.0 &5.073 &0.50 &0.0369 &204 & 135& 8553&62, 82\\

HB-Q3a5M1428 &HB &1.428 &1.590 &11.59 &0.182 &3.0 &4.284 &0.50 &0.0353 &171 & 136& 7182&62, 82\\

%\specialrule{0em}{1pt}{1pt}
15H-Q3a75M1691 &15H &1.691 &1.884 &13.72 &0.182 &3.0 &5.073 &0.75 &0.0367 &204 & 135& 8553&62, 82\\

HB-Q3a75M1428 &HB &1.428 &1.590 &11.59 &0.182 &3.0 &4.284 &0.75 &0.0351 &171 & 136& 7182&62, 82, 102\\

APR4-Q3a75M1366 &APR4 &1.366 &1.522 &11.08 &0.182 &3.0 &4.098 &0.75 &0.0358 &164 & 135& 6901&62, 82\\

H4-Q3a75M1651 &H4 &1.651 &1.836 &13.40 &0.182 &3.0 &4.950 &0.75 &0.0356 &199 & 135& 8347&62, 82\\

%\specialrule{0em}{1pt}{1pt}
15H-Q5a75M1691 &15H &1.691 &1.884 &13.72 &0.182 &5.0 &8.455 &0.75 &0.0411 &204 & 135& 8553&62, 82, 102\\

HB-Q5a75M1428 &HB &1.428 &1.590 &11.59 &0.182 &5.0 &7.140 &0.75 &0.0404 &171 & 136& 7182&62, 82, 102\\

\hline
H-Q3a5M1220 &H &1.220 &1.327 &12.26 &0.147 &3.0 &3.660 &0.50 &0.0351 &182 & 135& 7639&62, 82\\

B-Q3a5M1092 &B &1.092 &1.187 &10.97 &0.147 &3.0 &3.276 &0.50 &0.0357 &163 & 135& 6842&62, 82\\

%\specialrule{0em}{1pt}{1pt}
15H-Q3a75M1363 &15H &1.363 &1.484 &13.69 &0.147 &3.0 &4.089 &0.75 &0.0349 &203 & 135& 8523&62, 82\\

H-Q3a75M1220 &H &1.220 &1.327 &12.26 &0.147 &3.0 &3.660 &0.75 &0.0349 &182 & 135& 7639&62, 82, 102\\

B-Q3a75M1092 &B &1.092 &1.187 &10.97 &0.147 &3.0 &3.276 &0.75 &0.0356 &163 & 135& 6842&62, 82\\

APR4-Q3a75M1094 &APR4 &1.094 &1.190 &10.99 &0.147 &3.0 &3.282 &0.75 &0.0356 &163 & 135& 6842&62, 82\\

%\specialrule{0em}{1pt}{1pt}
H-Q5a75M1220 &H &1.220 &1.327 &12.26 &0.147 &5.0 &6.100 &0.75 &0.0449 &182 & 135& 7639&62, 82\\

B-Q5a75M1092 &B &1.092 &1.187 &10.97 &0.147 &5.0 &5.460 &0.75 &0.0451 &163 & 135& 6842&62, 82\\

\hline
\end{tabular}
}
\end{table*}

%% file: table_result_C18.tex
\begin{table*}[!htbp]
\caption{Characteristic physical quantities for the model with $C_{\rm{NS}}=0.182$. $\hat{M}_{\rm{rem}}$ is the normalized remnant mass for the matter located outside the apparent horizon, $\hat{M}_{\rm{eje}}$ is the ejecta mass during the merger, normalized by the baryon mass of the NS. $f_{\rm{cut}}m_0$ is the normalized cutoff frequency. $T_{\rm{eje}}$ and $v_{\rm{eje}}$ are the asymptotic kinetic energy and velocity of the ejecta. $\chi_{\rm{BH}}^f$ and $M_{\rm{BH}}^f$ are the dimensionless spin and mass of BHs after the merger. Models with EOS "B" are taken from Ref.~\cite{kyutoku2011sep}, with $C_{\rm{NS}}=0.1819$, $M_{\rm{NS}}=1.35M_{\odot}$, and $N=50$. Some values are N/A since Ref.~\cite{kyutoku2011sep} does not provide them. All physical quantities are evaluated at 10\,ms after the onset of merger. The grid resolution is $N=82$.} 
% $M_{\rm{fit}}^{2012}, M_{\rm{fit}}^{2018}$ are calculated from fitting formulas for the remnant mass given in Refs.~\cite{foucart2012dec,foucart2018oct}, and $M_{\rm{fit}}^{2016}, M_{\rm{fit}}^{2020}$ are calculated from fitting formulas for the ejecta mass given in Refs.~\cite{kawaguchi2016jun, krger2020feb}.
%\kenta{Kiuchi: In addition to aligning the digits, please fill out the blanks in the table. MS: It would be better to declare that $T_\mathrm{eje}$ in units of $10^{50}$\,erg is listed to erase e+49 and e+50. The same for the next two tables.}
%\kh{(comment: May be N62 result is not necessarily needed here, and the same for the other tables)}}
\label{tab:C18}
%\resizebox{\linewidth}{!}{
\begin{tabular}{ccc|ccccccc}
\hline
Model ID&&EOS&$\hat{M}_{\rm{rem}}$ %&$\hat{M}_{\rm{rem}}, N62$ 
&$\hat{M}_{\rm{eje}}$  %&$\hat{M}_{\rm{eje}}, N62$  
& $f_{\rm{cut}}m_0$ &$T_{\rm{eje}}[\rm{erg}]$ &$v_{\rm{eje}}[c]$ &$\chi_{\rm{BH}}^f$ &$M_{\rm{BH}}^f[M_{\odot}]$\\

\hline
%\multicolumn{12}{|c|}{Q3-a0.5}\\
%\hline
1&\multirow{2}{*}{Q3a5}&
15H &0.052%&0.0386 
& 0.0015%&0.00178
& 0.079&5.8$\times 10^{49}$ &0.15&0.77 &6.48\\

%15H-Q3a0.5N82 &0.0391 &0.0401 &-0.213 &0.0301 &1.82 &0.00133 &0.0 &0.443 &0.0 &0.443 &9.18e+49 &0.202 &0.237 &0.771 &6.5\\

2&&HB &0.047 %&0.0354 
&0.0015 %&0.00114
& 0.079&5.3$\times 10^{49}$ &0.16 &0.77 &5.48\\

&Ref.~\cite{kyutoku2011sep}&B&0.033&N/A&N/A&N/A&N/A&0.77&N/A\\

\hline
% \multicolumn{12}{|c|}{Q3-a0.75}\\
% \hline

%\multirow{4}{*}{15H}&15H&N62 &0.104 &0.127 &0.107 &0.00354 &0.0074 &0.00797 &2.27e+50 &0.195 &0.874 &6.42\\
3&\multirow{4}{*}{Q3a75}
&15H &0.12 %&0.104 
&0.0046 %&0.00354
& 0.071&2.3$\times 10^{50}$ &0.18 &0.87 &6.37\\

%\multirow{3}{*}{HB}&N62 &0.107 &0.127 &0.107 &0.00478 &0.00599 &0.00793 &2.40e+50 &0.188 &0.874 &5.41\\

4&&HB &0.12 %&0.107 
&0.0059 %&0.00485
& 0.069&2.8$\times 10^{50}$ &0.18 &0.87 &5.38\\

%&N102 &0.11 &0.127 &0.107 &0.00484 &0.00599 &0.00793 &2.31e+50 &0.183 &0.871 &5.4\\

%\multirow{2}{*}{APR4}&N62 &0.109 &0.126 &0.107 &0.00492 &0.00777 &0.00787 &2.52e+50 &0.194 &0.873 &5.17\\

5&&APR4 &0.12 %& 0.109 
&0.0065 %&0.00492
& 0.070&3.1$\times 10^{50}$ &0.19 &0.87 &5.15\\

%\multirow{2}{*}{H4}&N62 &0.11 &0.127 &0.107 &0.00349 &0.00435 &0.00795 &2.06e+50 &0.189 &0.874 &6.26\\

6&&H4 &0.13 %&0.110 
&0.0042 %& 0.00349
& 0.069&2.3$\times 10^{50}$ &0.18 &0.87 &6.22\\

&Ref.~\cite{kyutoku2011sep}&B&0.10&N/A&N/A&N/A&N/A&0.86&N/A\\

\hline
% \multicolumn{12}{|c|}{Q5-a0.75}\\
% \hline
7&\multirow{2}{*}{Q5a75}
&15H &0.037 %&0.0241 
&0.0047 %&0.00462
& 0.101&2.8$\times 10^{50}$ &0.19 &0.85 &9.81\\

%15H-Q5a0.75N82 &0.0273 &0.0278 &0.0 &0.0489 &0.0 &0.00635 &0.00392 &0.0 &0.00436 &0.0 &6.68e+50 &0.25 &0.267 &0.846 &9.83\\

%\multirow{3}{*}{HB}&N62 &0.0232 &0.0276 &0.0487 &0.00641 &0.0025 &0.0043 &5.83e+50 &0.253 &0.846 &8.31\\

8&&HB &0.036 %&0.0232 
&0.0054 %&0.00641 
& 0.101&2.8$\times 10^{50}$ &0.19 &0.85 &8.29\\

%&N102 &0.0252 &0.0276 &0.0487 &0.005 &0.0025 &0.0043 &4.60e+50 &0.254 &0.846 &8.3\\

&Ref.~\cite{kyutoku2011sep}&B&0.021&N/A&N/A&N/A&N/A&0.85&N/A\\

\hline
\end{tabular}
%}
\end{table*}

%% file: table_result_C147.tex
\begin{table*}[!htbp]
\caption{Characteristic physical quantities for the models with $C_{\rm{NS}}=0.147$. Models with EOS "H4" are taken from Ref.~\cite{kyutoku2015aug}, with $C_{\rm{NS}}=0.147$, $M_{\rm{NS}}=1.35M_{\odot}$, and $N=60$. $T_{\rm{eje}}$ and $v_{\rm{eje}}$ from Ref.~\cite{kyutoku2015aug} are not listed in the table since we use different definitions, which will be discussed in Appendix.~\ref{sec:appendix2}. $f_{\rm{cut}}m_0$ is also not listed since Ref.~\cite{kyutoku2015aug} does not provide it. Other quantities have the same definitions as in \autoref{tab:C18}.}
%\kenta{Kiuchi: In addition to aligning the digits, please fill out the blanks in the table.}
\label{tab:C147}
%\resizebox{\linewidth}{!}{
\begin{tabular}{ccc|ccccccc}
\hline
Model ID&&EOS&$\hat{M}_{\rm{rem}}$ %&$\hat{M}_{\rm{rem}}, N62$ 
&$\hat{M}_{\rm{eje}}$  
%&$\hat{M}_{\rm{eje}}, N62$  
& $f_{\rm{cut}}m_0$ &$T_{\rm{eje}}[\rm{erg}]$ &$v_{\rm{eje}}[c]$ &$\chi_{\rm{BH}}^f$ &$M_{\rm{BH}}^f[M_{\odot}]$\\

\hline
9&\multirow{2}{*}{Q3a5}
&H &0.16 %&0.140 
&0.024 %&0.0239
& 0.040&9.4$\times 10^{50}$ &0.18 &0.76 &4.59\\

10&&B &0.16 %&0.137 
&0.022 %&0.0215
& 0.041&8.4$\times 10^{50}$ &0.19 &0.76 &4.11\\

&Ref.~\cite{kyutoku2015aug}&H4&0.16&0.02&N/A&N/A&N/A&0.76&5.07\\

\hline
11&\multirow{4}{*}{Q3a75}

&15H &0.22 %&0.192 
&0.030 %&0.0311 
& 0.038&1.3$\times 10^{51}$ &0.18 &0.87 &5.07\\

%H-Q3a0.75N62 &0.193 &0.217 &-1.11 &0.21 &-0.782 &0.0292 &0.0314 &-0.213 &0.0319 &-0.261 &1.16e+51 &0.183 &0.86 &4.56\\

12&&H &0.21 %&0.193 
&0.030 %&0.0292 
& 0.039&1.2$\times 10^{51}$ &0.19 &0.87 &4.54\\

%B-Q3a0.75N62 &0.188 &0.217 &-1.38 &0.21 &-1.04 &0.0269 &0.0301 &-0.312 &0.0319 &-0.49 &1.10e+51 &0.196 &0.859 &4.1\\

13&&B &0.21 %&0.188 
&0.031 %&0.0269 
& 0.039&1.2$\times 10^{51}$ &0.19 &0.87 &4.07\\

14&&APR4 &0.21 %&0.189 
&0.031 %& 0.0313 
& 0.039&1.2$\times 10^{51}$ &0.19 &0.87 &4.07\\

&Ref.~\cite{kyutoku2015aug}&H4&0.22&0.03&N/A&N/A&N/A&0.88&4.99\\
\hline
15&\multirow{2}{*}{Q5a75}

&H &0.21 %& 0.181 
&0.040 %&0.0345 
& 0.059&2.2$\times 10^{51}$ &0.21 &0.83 &6.93\\

16&&B & 0.21%&0.178 
&0.039%&0.0347 
& 0.061&2.0$\times 10^{51}$ &0.22 &0.83 &6.21\\

%B-Q5a0.75N82 &0.189 &0.159 &1.59 &0.153 &1.87 &0.0357 &0.037 &-0.286 &0.0416 &-1.27 &1.84e+51 &0.221 &0.267 &0.832 &6.22\\

&Ref.~\cite{kyutoku2015aug}&H4&0.22&0.03&N/A&N/A&N/A&0.83&7.65\\

\hline
\end{tabular}
%}
\end{table*}

%% file: table_result_M18.tex
\begin{table*}[!htbp]
\caption{Characteristic physical quantities for the model with $M_{\rm{NS}}=1.8M_{\odot}$. Other quantities have the same definitions as in \autoref{tab:C18}. %\kenta{Kiuchi: Please use $\hat{M}_{\rm rem, eje}$ for the consistency.}
}
\label{tab:M18}
%\resizebox{\linewidth}{!}{
\begin{tabular}{cc|ccc|ccc|cccc}
\hline
&EOS &$\hat{M}_{\rm rem}$ &$\hat{M}_{\rm fit}^{2012}$ &$\hat{M}_{\rm fit}^{2018}$ &$\hat{M}_{\rm eje}$ &$\hat{M}_{\rm fit}^{2016}$ &$\hat{M}_{\rm fit}^{2020}$ &$T_{\rm eje} [\rm erg]$ &$v_{\rm eje}[c]$ &$\chi_{\rm BH}^f$ &$M_{\rm BH}^f[M_{\odot}]$\\

\hline
%15H-Q3a0.5N62 &0.194 &0.00665 &0.00268 &0.012 &0.000612 &0.0 &0.0 &5.28e+49 &0.218 &0.772 &6.96\\
\multirow{2}{*}{Q3a0.5}
&15H &0.014 &1.88$\times 10^{-3}$ &0.0117 &3.3$\times 10^{-4}$ &0 &0 &1.0$\times 10^{49}$ &0.13 &0.770 &6.95\\

&H &5.8$\times 10^{-5}$ &0 &0 &1.1$\times 10^{-6}$ &0 &0 &3.8$\times 10^{46}$ &0.13 &0.761 &6.93\\

\hline
%15H-Q3a0.75N62 &0.194 &0.0674 &0.0952 &0.0785 &0.00159 &0.00497 &0.00162 &1.17e+50 &0.202 &0.871 &6.87\\
\multirow{3}{*}{Q3a0.75}&
15H &0.081 &0.0945 &0.0779 &1.5$\times 10^{-3}$ &4.56$\times 10^{-3}$ &1.49$\times 10^{-3}$ &6.9$\times 10^{49}$ &0.16 &0.870 &6.83\\

%15H-Q3a0.75N102 &0.194 &0.0665 &0.0952 &0.0785 &0.00146 &0.00497 &0.00162 &9.90e+49 &0.194 &0.872 &6.85\\

%H-Q3a0.75N62 &0.219 &0.00357 &0.0291 &0.0314 &0.000297 &0.00584 &0.0 &2.83e+49 &0.227 &0.867 &6.91\\

&H  &8.9$\times 10^{-3}$ &0.0283 &0.0310 &1.1$\times 10^{-4}$ &5.48$\times 10^{-3}$ &0 &3.7$\times 10^{48}$ &0.14 &0.866 &6.90\\

%B-Q3a0.75N62 &0.252 &3.96e-05 &0.0 &0.000877 &1.74e-05 &0.019 &0.0 &2.40e+48 &0.271 &0.858 &6.88\\
&B &1.1$\times 10^{-5}$ &0 &7.73-04 &5.3$\times 10^{-7}$ &0.0187 &0 &2.3$\times 10^{46}$ &0.15 &0.858 &6.87\\

\hline
%15H-Q5a0.75N62 &0.194 &0.00278 &0.0 &0.0248  &0.000718 &0.0 &0.0 &8.02e+49 &0.248 &0.846 &10.5\\
\multirow{2}{*}{Q5a0.75}
&15H &6.3$\times 10^{-3}$ &0 &0.0244 &9.2$\times 10^{-4}$ &0 &0 &5.6$\times 10^{49}$ &0.18 &0.846 &10.48\\

%H-Q5a0.75N62 &0.219 &0.000117 &0.0 &2.69e-05 &5.64e-05 &0.0 &0.0 &5.21e+48 &0.224 &0.844 &10.5\\
&H &9.5$\times 10^{-6}$ &0 &6.58$\times 10^{-6}$ &2.6$\times 10^{-6}$ &0 &0&1.4$\times 10^{47}$ &0.17 &0.844 &10.47\\

\hline
\end{tabular}
%}
\end{table*}

%% file: table_chi_square.tex
\begin{table}[!htbp]
    \centering
        \caption{The $\chi$-squares for different fitting formulas. M18, C182, and C147 denote models with $M_{\rm{NS}}=1.8M_{\odot}, C_{\rm{NS}}=0.182$, and $C_{\rm{NS}}=0.147$, respectively. The first half is $\chi$-square calculated using $\Delta_{\rm{NR}}$ from Refs.~\cite{kawaguchi2016jun,foucart2012dec} to compare with the results in the previous simulations. The second half is $\chi$-square calculated using $\Delta_{\rm{NR}}$ obtained in our simulations.}
    \label{tab:chi_square}    
    \begin{tabular}{ccccc}
       \hline
       &$\chi_{\rm{rem}}^{2012}$ &$\chi_{\rm{rem}}^{2018}$&$\chi_{\rm{eje}}^{2016}$&$\chi_{\rm{eje}}^{2020}$\\
       \hline
       \multicolumn{5}{c}{$\Delta_{\rm{NR}}$ given in Eqs.~(\ref{eq:error esti kyutoku}),~(\ref{eq:error esti rem})}\\
%       \hline
       Models in Ref.~\cite{kyutoku2015aug}&4.75&5.56&0.0250&0.0601\\
       M18& 1.35& 2.75& 0.968& 0.00122\\
       C182&0.601& 3.38& 0.0320& 0.0300\\
       C147& 1.77& 5.47& 0.0499& 0.0162\\
       \hline
       \multicolumn{5}{c}{$\Delta_{\rm{NR}}$ given in Eq.~(\ref{eq:error esti})}\\
%       \hline
       M18&11.7& 28.6& 43.0& 0.0542\\
       C182&2.53& 10.9& 1.39& 1.31\\
       C147&2.15& 6.95& 1.20& 0.344\\
       \hline
    \end{tabular}
\end{table}

%% file: table_convergence.tex
\begin{table}[t]
    \caption{Several numerical results for models 15H-Q3a75M18,H-Q3a75M1220, HB-Q3a75M1428, and HB-Q5a75M1428 with different grid resolutions $N=62, 82$, and $102$. All the quantities are defined in the body text.}
    \label{tab:convergence}
    %\resizebox{\linewidth}{!}{
\begin{tabular}{cccccc}
    \hline
    N & $\hat{M}_{\rm rem}$& $\hat{M}_{\rm eje}$& $v_{\rm{eje}}$ &$ f_{\rm{cut}}m_0$& $\chi_{\rm{BH}}$\\
    \hline
    &15H-Q3a75M18&&&&\\
    62& 0.0840& 0.00144& 0.155& 0.0762& 0.870\\
    82& 0.0807& 0.00151&  0.159 & 0.0849& 0.870\\
    102 & 0.0782& 0.00203& 0.172& 0.0883& 0.870\\ \hline
    &H-Q3a75M1220&&&&\\
    62& 0.213& 0.0290& 0.181& 0.0384& 0.863\\
    82& 0.212& 0.0296& 0.185& 0.0394& 0.871\\
    102 &0.210& 0.0303& 0.188& 0.0398& 0.868\\ \hline
    &HB-Q3a75M1428&&&&\\
    62& 0.128& 0.00513& 0.172& 0.0666& 0.873\\
    82& 0.124& 0.00589& 0.182& 0.0687& 0.870\\
    102 &0.122& 0.005884& 0.185& 0.0701& 0.870\\ \hline
    &HB-Q5a75M1428&&&&\\
    62& 0.0393& 0.000414& 0.200& 0.0993& 0.845\\
    82& 0.0360& 0.00542& 0.191& 0.1014& 0.845\\
    102 & 0.0327& 0.00527& 0.192& 0.1016& 0.845\\ \hline
    &15H-Q5a75M1691&&&&\\
    62& 0.0428& 0.000463& 0.177& 0.0993& 0.845\\
    82& 0.0369& 0.00474& 0.187& 0.1011& 0.845\\
    102 & 0.0351& 0.00491& 0.185& 0.1016& 0.845\\
    \hline
\end{tabular}
    %}
\end{table}